\documentclass[aps,pre,superscriptaddress,showpacs]{revtex4}

\usepackage{graphicx}
\usepackage{bm}

\begin{document}
\title{Continuum-particle hybrid coupling for
 mass, momentum and energy transfers in unsteady fluid flow}

\author{R. ~Delgado-Buscalioni}
\email[]{R.Delgado-Buscalioni@ucl.ac.uk} \affiliation{Centre for
Computational Science, Dept. Chemistry, University College London, London,
U.K} \author{P. V. Coveney} \email[]{P.V.Coveney@ucl.ac.uk}
\affiliation{Centre for Computational Science, Dept. Chemistry, 
University College London, London, U.K}

\begin{abstract}
The aim of hybrid methods in simulations is to communicate regions
with disparate time and length scales. Here, a fluid described at the
atomistic level within an inner region P is coupled to an outer region
C described by continuum fluid dynamics.  The matching of both
descriptions of matter is made across an overlapping region and, in
general, consists of a two-way coupling scheme (C$\rightarrow$P and
P$\rightarrow$C) which conveys mass, momentum and energy fluxes.  The
contribution of the hybrid scheme hereby presented is two-fold: first
it treats unsteady flows and, more importantly, it handles energy
exchange between both C and P regions.  The implementation of the
C$\rightarrow$P coupling is tested here using steady and unsteady
flows with different rates of mass, momentum and energy exchange. In
particular, relaxing flows described by linear hydrodynamics
(transversal and longitudinal waves) are most enlightening as they
comprise the whole set of hydrodynamic modes.  Applying the hybrid
coupling scheme after the onset of an initial perturbation, the
cell-averaged Fourier components of the flow variables in the P region
(velocity, density, internal energy, temperature and pressure) evolve
in excellent agreement with the hydrodynamic trends.  It is also shown
that the scheme preserves the correct rate of entropy production. We
discuss some general requirements on the coarse-grained length
and time scales arising from both the characteristic microscopic and
hydrodynamic scales.
\end{abstract}


\maketitle

\section{Introduction}

A  wide range of  systems with  important applications are  governed  by a fine  interplay
between the atomistic processes occurring within a small region of the system and the slow
dynamics occurring within the bulk. A  large list of examples arise  in complex flows near
interfaces (polymers or colloids near surfaces,  wetting, drop formation, melting, crystal
growth from a  fluid phase, moving  interfaces of immiscible fluids  or membranes, to name
only a few).    The  computational expense of  realistic-size  simulations  of  these
problems via standard molecular dynamics (MD) is prohibitive, and such kind of studies
require new algorithms which can retain the benefit of the atomistic description of matter
where it is  really  needed, while  treating the  bulk of  the system by  much less costly
continuum fluid mechanics methods.

Several hybrid algorithms of this sort have been proposed in the recent literature.  
In general, to couple the particle region P and the continuum region C,
such hybrid schemes use an overlapping region comprised of two buffers
 C$ \rightarrow $P and P$\rightarrow$C,
which account for the two-way transfer of information: from C to P and
{\em vice versa} (see Fig. 1). While the P$\rightarrow$C transfer essentially consists of
a coarse-graining procedure,  at C$ \rightarrow $P one needs to 
reconstruct the dynamics of a large collection of particles with only
the limited prescription from the C region as input. 
Moreover, in performing this reconstruction, 
the number of unphysical artifacts added
(as Maxwell d{\ae}mons) should be minimized far as possible.
This task is very complicated and represents in fact
the heart of any hybrid scheme.

Hybrid algorithms for fluids are relatively recent.  The elegant
method introduced by Garcia {\em et al.}  \cite{Gar00} for rarefied
gases couples fluxes arising from a direct simulation Monte Carlo
(DSMC) scheme to another region described by computational fluid
dynamics (CFD). The DSMC is set at the finest grid scale of an
adaptive mesh refinement hierarchy, while a CFD semi-implicit solver
is used at the upper level scales.  In passing, we note that the
scheme may, in principle, be implemented using an (MD-Continuum)
liquid description, although in this case the C-solver must be
completely explicit to avoid having to change the particle's energy in
the iterations of the implicit scheme.

In the case of liquids, the state of the art is relatively less
developed due to the complications arising from the interparticle
forces.  A pioneering work by O'Connel and Thompson \cite{Tho95}
coupled momentum by imposing the local continuum velocity at
C$\rightarrow$P via a crude constraint Lagrangian dynamics.
Hadjiconstantinou and Patera \cite{Had97} introduced a reservoir
region to impose boundary conditions on the P region (this reservoir
being the equivalent of the C$\rightarrow$P domain defined
here). While in residence in the reservoir, particles were given at
each time step, a velocity drawn from a Maxwellian distribution with
mean and variance consistent with the velocity and temperature of the
C-flow.  To obtain the boundary condition for the C region the authors
used a low order polynomial to smooth the field variables derived from
P at the P$\rightarrow$C region.  In order to match the boundary
conditions for both the P and C regions, Hadjiconstantinou and Patera
\cite{Had97} implemented an iterative scheme (based on the Schwarz
alternating method) that is suitable for steady incompressible flows.
Liao and Yip \cite{Liao98} proposed a sophisticated method (called the
thermodynamic field estimator) to extract continuum fields from the
particle data by means of maximum likehood inference.  This idea may
be used to ameliorate the P$\rightarrow$C coupling when the flow
presents large gradients, albeit at a rather large computational
cost. To transfer momentum on the P region 
Liao and Yip \cite{Liao98} proposed a new Maxwell d{\ae}mon, called
reflecting particle method. A drawback is that the pressure gradient
is then an outcome of the simulation, rather than an input.  Finally,
Flekkoy {\em et al.}  \cite{Flek00} used the idea of
coupling-through-fluxes and also implemented mass transfer.  However,
energy transfer was still not allowed and only steady flows were
considered.  The main purpose of the present work is to broaden the
scope of such hybrid schemes towards a general description allowing
mass, momentum and energy coupling in unsteady flows.

A question of central interest is to decide what kind of information
needs to be transfered at C$ \rightarrow $P and P$\rightarrow$C.
There are essentially two possibilities, to transfer either
generalised forces (fluxes of conserved quantities) or the local
values of the averaged-variables.  Both kinds of approaches can be
found in the published literature.  Here, in the context of energy
transfer, we show that under unsteady flows it is not sufficient to
impose the local C-quantities at the boundary of P; instead, it is
necessary to couple through fluxes. Another possible benefit of
flux-coupling was pointed out by Flekkoy {\em et al.}  \cite{Flek00}
who stated that this procedure transcends the problem of working with
fluids whose constitutive relations may be only partially or
incompletely known.  Although we agree that the flux-based coupling is
the correct matching procedure, we show nevertheless that if the
transport coefficients at C and P are disparate enough, the hybrid
scheme fails to couple the time evolution of both domains. Hence, in
such cases, the evaluation of transport coefficients (using standard
microscopic techniques, at least for the range of densities and
temperatures under study) is an unavoidable requirement for the
correct behaviour of the hybrid scheme.

The rest of the paper proceeds as follows.  The equations governing
the continuum and particle regions and the averaging procedures are
presented in Sec. \ref{2}. The core of the scheme, describing the
C$\rightarrow$P coupling for momentum, energy and mass fluxes, is
presented in Sec \ref{3}. General requirements on the coarse-graining
length and time scales are discussed in Section \ref{req}. The
unsteady flows under which the scheme has been tested (decay of
longitudinal and transversal waves) are presented in Sec. \ref{hyd}
and in Sec. \ref{4} we discuss the results of these tests as far as
the main hydrodynamic and thermodynamic variables are
concerned. Finally Sec. \ref{con} is devoted to conclusions.

\section{\label{2}
Overview and geometry of the hybrid coupling scheme}

The domain decomposition of the hybrid scheme is depicted in Fig. 1.
Two regions need to be distinguished: the particle region P, and the
continuum region C. Region P is composed of an ensemble of particles
interacting through prescribed interparticle potentials and evolving
in time through Newtonian dynamics. In order to illustrate the
coupling procedure a Lennard-Jones (LJ) fluid will be considered.
Within P a number $N(t)$ of particles, located at $\bf{r}=
\left\{\bf{r}_i\right\}$ (the subscript $i$ denoting the $i^{th}$
particle) interacts through the LJ potential
$\psi(r)=4\epsilon^{-1}\,\left[(\sigma/r)^{12}-(\sigma/r)^{6}\right]$.
Each particle has a mass $m$, velocity $\bf{v}_i$ and energy
$\epsilon_i= \frac12 m v_i^2+ \Sigma_{j} \psi(r_{ij})$
($\mathbf{r}_{ij}=\mathbf{r}_j-\mathbf{r}_i$).  Their equations of
motion,
\begin{eqnarray}
\dot\mathbf{r}_i&=&\mathbf{v}_i,\\
\dot\mathbf{v}_i&=&\mathbf{f}_i/m=\Sigma_{j=1}^{N} \frac{d \psi(r_{ij})}{d r_{ij}} 
\frac{\mathbf{r}_{ij}}{r_{ij}},
\end{eqnarray}
are solved via standard molecular dynamics (MD) at time steps $\Delta
t_P\simeq 10^{-3} \tau$, where $\tau=(m\sigma^2/\epsilon)^{1/2}$ is
the characteristic time of the LJ potential.  Throughout the rest of
the paper, all quantities will be expressed in reduced units of the LJ
potential: $\tau(=0.45\times 10^{-13}$s), $\sigma(=3.305\times
10^{-12}$cm), $\epsilon$, $m(=6.63\times 10^{-23}$g) and
$\epsilon/k_B(=119.18$K) for time, length, energy, mass and
temperature , respectively (the numerical values correspond to argon).

On the other hand, within the C region the relevant variables are the
 macroscopic local densities associated with the conserved quantities,
 the number density $\rho(\bf{ R},t)$, the energy density $\rho e(\bf{
 R},t)$ and the momentum density $\bf{j}(\bf{ R},t)$ (related to the
 local mean velocity $\bf{ u}$ by $\bf{j}=\rho\bf{u}$). In what
 follows the spatial coordinates of the macroscopic fields are denoted
 by capital letters $\mathbf{R}$, while the the microscopic
 coordinates are designated by lower case letters.  The conservation
 laws for the local densities are
\begin{eqnarray}
\label{mass}
\frac{\partial \rho}{\partial t}&=& -\nabla \cdot \rho \mathbf{u} ,\\
\label{mom}
\frac{\partial \mathbf{j} }{\partial t}&=& -\nabla\cdot\left(\mathbf{j u} + \bm{\Pi}\right),\\
\label{ener}
\frac{\partial \rho e}{\partial t}&=& -\nabla\cdot\left(\rho e \mathbf{u} + \bm{\Pi}\cdot \mathbf{u} + \mathbf{q} \right),
\end{eqnarray}
where the specific energy $e=u^2/2 + 3T/2 + \phi$, includes the translational energy,
the thermal kinetic energy and the potential energy $\phi$. The momentum flow 
contains contributions from convection $\mathbf{j u}$ and the pressure tensor
$\bm{\Pi}= P\,\mathbf{1} + \bm{\tau}$, the latter including the local hydrostatic pressure 
$P(\bf{R},t)$ and the viscous stress tensor, which 
satisfy a Newtonian constitutive relation,
as shown by previous previous MD descriptions of the  LJ-fluid \cite{TRANSP,Ciccotti},
\begin{equation}
\label{tau}
\bm{\tau}= -\eta \left(\nabla\mathbf{ u} + (\nabla \mathbf{u})^{T}- \frac{2}{3}\nabla\cdot\mathbf{ u}\right) -  \xi \nabla\cdot\mathbf{ u}.
\end{equation}
The energy current includes convection $\rho e \mathbf{u}$,
dissipation $\bm{\Pi}\cdot \mathbf{u}$ and conduction $\mathbf{q}$, 
which can be expressed in terms
of the local temperature gradients 
and the thermal conductivity $\kappa_c$, through Fourier's law 
$\mathbf{q}=-\kappa_c \nabla T(\mathbf{R},t)$.
In order to close the above equations it is necessary to know the caloric $e(\rho,T)$
and thermal equations of state $P(\rho,T)$ and the constitutive relations for
the transport coefficients (shear and bulk viscosities and thermal conductivity; $\eta$,
$\xi$ and $\kappa_c$ respectively) in terms of a set of independent thermodynamic variables,
such as $\rho$ and $T$.  The equations of state for a LJ-fluid were extracted 
from Johnson {\em et al.} \cite{EOS} and 
the transport coefficients $\eta$, $\kappa_c$ and $\xi$ from Heyes \cite{TRANSP}
and Borgelt {\em et al.} \cite{BulkVis}.
The variables relevant to the C region are the slower ones.
Using any standard continuum fluid dynamics solver (e.g. based on a finite volume method)
the evolution of the C-variables will be traced at time intervals $\Delta t_C>>\Delta t_P$
and evaluated within cells of volume $V_l$
whose size and location is given by the nodes of a
certain mesh, $\left\{\mathbf{R}_l\right\}$, $l=\left\{ 1,...,M_c \right\}$.
It will be assumed that the size of the C$\rightarrow$P and P$\rightarrow$C regions 
are the same size as those of the cells used in the spatial discretisation of the
selected continuum solver, say $V_l=(\Delta X)^3$.
In general both $\Delta X$ and $\Delta t_C$
may depend on the type of solver used for the C-region, or
on the characteristic length of the particular phenomena under study,
Nevertheless, various intrinsic constraints on $\Delta X$ and $\Delta t_C$
will be mentioned in Sec.\ref{req}.

\subsubsection{Averages}
Averages are needed in order to transfer information from the faster
time and shorter length-scale particle dynamics to the slower and
longer coarse-grained description.  In order to deal with unsteady,
non-equilibrium scenarios, averages need to be local on the slower
time scale and in the coarse-grained spatial coordinates.  For any
particle variable, say $\Phi_i$, we define the following averages:
\begin{eqnarray}
\label{sum}
\bar{\Phi}(\mathbf{R}_l,t) &\equiv & \frac{1}{N_l}\sum_{i \in V_l}^{N_l} \Phi_i,\\
\left< \bar{\Phi}\right>(\mathbf{R}_l,t_C)&\equiv & \frac{1}{\Delta t_C}\int_{t_C}^{t_C+\Delta t_C} \bar{\Phi}(\mathbf{R}_l,t)dt
\end{eqnarray}
where the summation  in Eq. (\ref{sum}) is made over the $N_l$ particles inside the cell $l$.

The averaging procedure is needed to translate the P and C
``languages'' to and from each domain.  This translation is done
within the overlapping region, where the two descriptions of matter
coexist (see Fig. 1). In particular, within the P$\rightarrow$C cells
the many degrees of freedom arising from the particle dynamics are
coarse-grained to provide boundary conditions at the ``upper''
C-level.  As long as the number of degrees of freedom is very much
larger at P than at C, this operation is rather straightforward and is
based on the microscopic derivation of continuum fluid dynamics
\cite{Evans}.  We adopt the approach advocated by Flekkoy {\em et al.}
\cite{Flek00}, in making the information transferred from P to C to be
the coarse-grained particle-fluxes of conserved quantities. These are

\begin{eqnarray}
\rho\mathbf{u}\cdot\mathbf{n}_{PC}&=&\frac1V_{PC} \left<\Sigma_{i=1}^{{N}_{PC}} m \mathbf{v}_i\right>\cdot\mathbf{n}_{PC} \\
\bm{\Pi}\cdot\mathbf{n}_{PC}&=&\frac1V_{PC} \left< \left( \Sigma_{i=1}^{{N}_{PC}} m \mathbf{v}_i\mathbf{v}_i -\frac12\Sigma_{i,j}^{{N}_{PC}} \mathbf{r}_{ij}\mathbf{F}_{ij}\right)\right>\cdot\mathbf{n}_{PC}\\
\mathbf{q}\cdot\mathbf{n}_{PC}&=&\frac1V_{PC} \left< \left(\Sigma_{i=1}^{{N}_{PC}} m \epsilon_i\mathbf{v}_i  -\frac12\Sigma_{i,j}^{{N}_{PC}} \mathbf{r}_{ij} 
\mathbf{v}_i\mathbf{F}_{ij}\right)\right>\cdot\mathbf{n}_{PC}
\end{eqnarray}
where $N_{PC}$ is the number of particles inside the P$\rightarrow$C cell and $\mathbf{n}_{PC}$ 
is the surface vector shown in Fig. 1.

By contrast, within the C$\rightarrow$ P cells, the particle dynamics
must be modified to conform to the averaged-dynamics prescribed by the
continuum description.  In other words, one needs to construct a sort
of ``generalised boundary condition'' for the particle dynamics. As
pointed out in all previous papers on the subject
\cite{Gar00,Tho95,Had97,Flek00}, this represents the most demanding
challenge in that one needs to {\it invent} a way to reconstruct the
microscopic dynamics of a large number of particles, based on only a
few properties of the local continuum variables. Moreover, to ensure
that the effect on the inner P-region is minimised, it is crucial to
reduce as much as possible the unphysical artifacts, such as Maxwell
d{\ae}mons, that are added to the particle dynamics at
C$\rightarrow$P.  The present work is focused on this problem, which
lies at the heart of any hybrid scheme.

\section{\label{3}
The C$\rightarrow$P coupling} This part of the hybrid scheme can be
alternatively stated as the imposition of generalised (mass, momentum
and energy) boundary conditions on an MD simulation box.  To deal with
this task we have coupled the particle region to a collection of flows
(with explicit analytical solution), which involve the whole set of
conserved quantities exchanged (mass, momentum and energy). In this
sense, in the present work our C-solver is not numerical but rather
analytical. In particular, we use the initial
(non-equilibrium) state imposed at P to calculate the 
time-dependent analytical solution at C. 
This C-flow is then imposed on the P region
during the rest of the simulation, meaning that (appart from the
initial state) the hybrid coupling used in the tests presented here
works in one direction only (from C to P).

\subsection{\label{uns}
Imposing fluxes under unsteady flows} Following Flekkoy {\em et al.}
\cite{Flek00}, at C$\rightarrow$ P we shall communicate fluxes of
conserved quantities.  These fluxes correspond to mass, momentum and
energy transfers through the outer interface of the C$\rightarrow$ P
cell (the W-surface in Fig. 1).  Flekkoy {\em et al.} \cite{Flek00}
obtained these fluxes from the values of the continuum variables at
the centre of the control cell, $x=x_O$, instead of at the exact
position of the C$\rightarrow$P interface, $x=x_W$. We have found that
it is essential to take into account this apparently unimportant
technicality when dealing with unsteady scenarios.  Let us consider a
general conservation equation, with
\begin{equation}
\label{cons}
\frac{\partial \phi}{\partial t}+ \nabla \cdot \mathbf {J_{\phi}} =S_{\phi},
\end{equation}
where $J_{\phi}$ is the flux of $\phi$ and the source term vanishes,
$S_{\phi}=0$, as in Eqs. (\ref{mass})-(\ref{ener}).  Integrating over
the control cell C$\rightarrow$P,
\begin{equation}
\frac{\partial}{\partial t} \int_{\mathit V}  \phi dV + \int_{\mathit S} \mathbf{ J_{\phi}} \cdot \mathbf{n} ds = 0.
\end{equation}
For illustration we shall restrict analysis to the one-dimensional
(1D) situation depicted in Fig. 1. In this case one obtains,
\begin{equation}
\label{int.vol}
\frac{\partial}{\partial t} \int_{\mathit V} \phi dV  - A\mathbf{J_{\phi}}_E \cdot \mathbf{n}_W= -A \mathbf{J_{\phi}}_W \cdot \mathbf{n}_W,
\end{equation}
where the subscripts $E$ (east) and $W$ (west) denote that the
variables are measured at $x=x_E$ and $x=x_W$ respectively.  The
surface vectors, $\mathbf{n}_W$ and $\mathbf{n}_E$ are shown in
Fig. 1, and in Eq. (\ref{int.vol}) use has been made of
$\mathbf{n}_W=-\mathbf{n}_E$.  The R.H.S. of Eq. (\ref{int.vol}) is
the flux current of $\phi$ through the interface $W$ of the control
cell, which is precisely the (generalised) force we want to introduce
on the particles at the C$\rightarrow$P buffer.  We note that only
under steady flows does $\mathbf{J_{\phi}}_W=\mathbf{J_{\phi}}_O$ (to
see this, integrate Eq.  (\ref{cons}) from $x=x_O$ to $x=x_W$): hence
only in this case does the evaluation of the fluxes at $x_W$ using the
continuum variables at $x_O$ lead to the same converged steady state
as if the variables at $x_W$ were used (although the transients may of
course differ). It is possible to provide an estimate of the global
error arising from evaluating the flux at a position $x_O$ shifted
$\delta\, \Delta X$ with respect $x_W$, over a certain time interval
$\Delta t$.  In the case of the momentum equation the deviation of the
stress contribution to the momentum flux $J=\mathbf{J}\cdot\mathbf{n}$
at any instant would be of order $\Delta J\simeq \nabla J\,
\delta\Delta X$, with $\delta =| x_O-x_W|/\Delta X$ is the distance to
the C$\rightarrow$P interface ($\delta =0.5$ in Fig. 1). Assuming that
the mean velocity field can be expressed as
$u=\mathbf{u}\cdot\mathbf{n} \sim u^{(k)}\exp(i k x)$ ($k$ being the
dominant wavenumber) and using the Newtonian constitutive relation for
the viscous tensor in Eq. (\ref{tau}), one obtains $\Delta J \sim
\eta_L k^2 u^{(k)} \delta \Delta X$, where $\eta_L=4\eta/3+\xi$ is the
longitudinal viscosity.  As a particular example we consider a
longitudinal wave and evaluate the error along a cycle, $\Delta t=2
\pi/(kc_s)$ (where $c_s$ is the sound velocity).  As a crude estimate,
the accumulated error of the cell-averaged momentum
$\mathbf{j}\cdot\mathbf{n}=\rho_e\bar{u}$ is of order $\rho_e \Delta
\bar{u}\sim \Delta J \Delta t/\Delta X$; using $\rho_e=0.5$,
$\eta_L\simeq 1 $ and $c_s\simeq 5$, one obtains $\Delta
\bar{u}/u^{(k)}\sim 2\pi \delta \eta_L k/(c_s \rho_e) \simeq 0.3$, for
the typical wavenumbers considered here $k\sim 0.2$.  Simulations
carried out with the momentum flux evaluated at $x_O$ yield relative
errors of the averaged velocity at C$\rightarrow$P of the same order
of magnitude as this estimate (see Sec \ref{momt} and Fig. 6b).  We
observe that most computational fluid dynamics (CFD) codes provide the
continuum variables at the centre of the control cells (i.e., at
$x_O$), so that in order to evaluate the fluxes pertaining to the
C$\rightarrow$P exchange it would first be necessary to make use of
simple interpolation techniques.

The fluxes arising from the continuum equations (on the R.H.S. of
Eqs. (\ref{flux.mass}-\ref{flux.energ})) are imposed on the particle
ensemble at the C$\rightarrow$P cells through expressions involving
atomistic variables (those on the R.H.S. of
(\ref{flux.mass}-\ref{flux.energ})).

\begin{eqnarray}
\label{flux.mass} m\mathrm{s} &=&-A \rho\mathbf{u} \cdot \mathbf{n},\\
\label{flux.mom} m\mathrm{s}\left<\mathbf{v^{\prime}}\right> + \left<\sum_i^{N_{CP}}
\mathbf{F}_i^{ext}\right>&=&-A\left(\rho \mathbf{u u} + {\bm \Pi}
\right)\cdot {\mathbf n}, \\
\label{flux.energ} 
m\mathrm{s}\left<\epsilon^{\prime}\right>  +\left<\sum_{i}^{N_{CP}} \mathbf{F}_i^{ext}\cdot \mathbf {v}_i\right> -\left<\mathbf{J}_Q^{ext}\right>\cdot \mathbf{n} &=&-A\left(\rho {\mathbf u}\, e + \bm{\Pi\cdot u} +\mathbf{q}\right)\cdot \mathbf{n}.
\end{eqnarray}
where henceforth, $\mathbf{n}$ indicates the vector of the outermost
interface of the C$\rightarrow$P cell, pointing towards C.  The
nomenclature used here follows that of Flekkoy {\em et al.}
\cite{Flek00}: $\mathrm{s}(t)$ indicates the number of particles
inserted ($\mathrm{s}>0$) or removed ($\mathrm{s}<0$) from
C$\rightarrow$P per unit of time; the velocity of the inserted/removed
particles is $\mathbf{v^{\prime}}$; $\mathbf{F}_i^{ext}$ is the
external force applied to each particle $i$ within the C$\rightarrow$P
cell. The total external force is $\Sigma^{N_{CP}}
\mathbf{F}_i^{ext}$, where the summation is over the $N_{CP} (t)$
particles inside C$\rightarrow$P. Finally, $\left<
\epsilon^{\prime}\right>$ indicates the energy of the inserted/removed
particles and $\left<\mathbf{J}_Q^{ext}\right>$ refers to an
externally imposed heat current.

As mentioned by Flekkoy {\em et al.} \cite{Flek00}, insertion of
Eq. (\ref{flux.mass}) into Eq.(\ref{flux.mom}) shows that the rates of
change of momentum due to convection and local stresses are correctly
introduced if $\left<\mathbf{v}^{\prime}\right>=\mathbf{u}$ and
$\left<\sum_{i}^{N_{CP}} \mathbf{F}_i^{ext}\right> - =A\,{\bm
\Pi}\cdot \mathbf {n} =-A\,\left (P \mathbf {n}+ \bm{\tau}\cdot\mathbf
{n}\right)$, respectively.  Note that $-P\mathbf {n}$ is the
hydrostatic pressure force (pointing inwards the C$\rightarrow$P
cell), while the viscous contribution $-\bm{\tau}\cdot\mathbf {n}$
depends on the local velocity gradient.

The balance of the energy flux requires some extra conditions. In
Eq. (\ref{flux.energ}) the convection, dissipation and conduction of
energy are balanced if $\left<\epsilon^{\prime}\right>=e$,
$\left<\sum_i^{N} \mathbf{F}_i^{ext}\cdot \mathbf{v}_i\right>=
-A\,\bm{\Pi}\cdot\mathbf{u}\cdot \mathbf{n}$ and
$\left<\mathbf{J}_Q^{ext}\cdot\mathbf{n}\right> =A\,\mathbf{q}\cdot
\mathbf{n}$ respectively .

Let us now consider in more detail how the scheme deals with momentum,
energy and mass transfer from C to P.

\subsection{Momentum exchange}

The condition $\left<\mathbf{v}^{\prime}\right>=\mathbf{u}$ ensures
the balance of momentum convection. If the mass flux points towards
the P region ($s>0$), this condition is fulfilled by choosing the
velocity of the inserted particles from a Maxwellian distribution
according to the local temperature at the C$\rightarrow$P cell,
$P(\mathbf{v}^{\prime})=(1/2 \pi
mkT)^{3/2}\exp\left(-m(\mathbf{v}-\mathbf{u})^2/2mkT\right)$.
Concerning particle removal $(s<0)$, we note that if the average
velocity at the C$\rightarrow$P cell is equal to the continuum
velocity $\left<\mathbf{\bar{v}}\right>=\mathbf{u}$, then the average
velocity of the subset of extracted particles would be precisely
$\left<\mathbf{v}^{\prime}\right>=\left<\mathbf{\bar{v}}\right>=\mathbf{u}$. Hence
the condition of velocity continuity at C$\rightarrow$P is needed to
ensure the correct balance of momentum convection.

The change of momentum due to local stresses establishes the overall
external force exerted on the particle region. Therefore one needs to
determine how the overall external force is distributed on each
individual particle.  Flekkoy {\em et al.} \cite{Flek00} distributed
the force according to a certain function $g(x)$ satisfying
$g(x_W)=\infty$, $g(x_O)=g^{\prime}(x_O)=0$. Normalisation leads to
\begin{equation}
\mathbf{F}_i^{ext}= -\left(\frac{g(x_i)}{\sum_{i}^{N_{CP}} g(x_i)}\right)\;A {\bm \Pi}\cdot \mathbf {n} ,
\end{equation}
where $x$ runs perpendicular to the C$\rightarrow$P interface and the
applied force is made constant along each $\Delta t_C$.  As $g(x)$
tends to infinity as $x\rightarrow x_W$, the applied force diverges as
one approaches the C$\rightarrow$P interface; hence the density nearby
$x=x_W$ is very small or zero.  The function $g(x)$ is thus endowed
with a two-fold purpose: it ensures a limiting extension to P (as the
hydrostatic pressure force always points towards the P region,
particles will never cross the C$\rightarrow$P interface outwards)
while also guaranteeing the existence of a small region where
particles can be inserted with very low risk of overlapping.

Despite the benefits of the $g(x)$ function for distributing the
externally imposed momentum, we decided to use $g(x)=1$ for all $x$
inside the C$\rightarrow$P cell. The reasons for this choice will
become clear when explaining the energy exchange, below. The first
implication of $g(x)=1$ is that the external force is equally
distributed among all the particles within the C$\rightarrow$P
cell. In other words, $\mathbf{F}_i^{ext}$ no longer depends on the
particle label
\begin{equation}
\mathbf{F}_i^{ext}=\mathbf{F}^{ext}=-\left(\frac{1}{N_{CP}}\right)\;A {\bm \Pi}\cdot \mathbf {n}.
\end{equation}

The second implication of $g(x)=1$ is that the particle density
profile near the C$\rightarrow$P interface no longer vanishes, so one
needs an efficient way to resolve the problem of overlap on the
insertion of new particles. This task is carried out by the {\sc
usher} algorithm, as explained below. Finally in order to ensure a
finite extent of the particle region, if a particle ($i$) crosses
outwards the C$\rightarrow$P interface (in Fig. 1, $x_i=x_W-\delta$,
with $\delta>0$) with velocity $\mathbf{v}_i$ it is substituted by
another one ($j$) with $y_j=y_i;\,z_j=z_i$, $x_j=x_W+\delta$ and with
$\mathbf{v}_j =\mathbf{v}_i$. In this way, the overall momentum is
strictly conserved before and after the particle exchange.

\subsection{Energy exchange
\label{enerbal}}

\subsubsection{Advection}
The balance of advected energy requires that
$\left<\epsilon^{\prime}\right>= e=u^2/2 + 3T/2 +\phi$.  Decomposing
the particle energy into the kinetic and potential parts
$\epsilon^{\prime}=[v^{\prime}]^2/2 + \psi^{\prime}$, one sees that,
since the new particles are drawn from a Maxwell distribution,
$\left<[v^{\prime}]^2/2\right>=u^2/2 + 3T/2$. By contrast, the balance
of potential energy $\left<\psi^{\prime}\right>=\phi$ is much less
straightforward to implement.  This condition is fulfilled by the {\sc
usher} algorithm, described below.

\subsubsection{Dissipation}
One needs to satisfy the following balance of heat dissipation:
\begin{equation}
\label{dis}
\left<\sum_i^{N_{CP}} \mathbf{F}_i^{ext}\cdot \mathbf{v}_i\right>=-A\,\bm{\Pi} \cdot\mathbf{u}\cdot \mathbf{n}.
\end{equation}
This condition does not generally hold if the external force is
distributed according to an arbitrary $g(x)$.  Indeed it is not even
clear that a function $g(x)$ exists satisfying $g(x_W)\rightarrow
\infty$ and enabling the heat dissipation balance in Eq. (\ref{dis}).
In any case, such a function $g$ would depend on the particle's
velocity distribution and then the problem of finding $g$ would become
a formidable task at each time step.

The advantage of using $g(x)=1$ now becomes clear.  As long as
$\mathbf{F}_i^{ext}$ does not depend on the particle label, one can
greatly simplify the L.H.S. of Eq. (\ref{dis}) equation to obtain,
\begin{equation}
\label{dis2}
\mathbf{F}^{ext} \cdot \left<\sum_i^{N_{CP}} \mathbf{v}_i\right>=
N_{CP} \mathbf{F}^{ext} \cdot \left< \mathbf{\bar{v}}\right>
= -A\,\bm{\Pi} \cdot\mathbf{u}\cdot \mathbf{n}.
\end{equation}
The last equality follows from construction of the overall force
$N_{CP} \mathbf{F}^{ext}= -A\bm{\Pi}\cdot \mathbf{n}$ and from the
continuity of velocity $\left< \mathbf{\bar{v}}\right>=\mathbf{u}$.

\subsubsection{\label{cond} Conduction}

The condition $\left<\mathbf{J}_Q^{ext}\right>\cdot\mathbf{n} =A\,\mathbf{q}\cdot
\mathbf{n}$ requires the establishment of a heat current along the
C$\rightarrow$P cell representing the conduction of energy. This may be
implemented by various means; for instance, following the idea of
Evans and coworkers \cite{Evans} one may include an extra force which
pulls the ``hotter'' particles towards the direction of the heat flux 
and conserves the overall momentum. Alternatively, one may try to
impose a Chapman-Enskog velocity distribution with the desired heat
flux, at some region inside the C$\rightarrow$P buffer. In this work we have 
made use of the phenomenological Fourier's law, $\mathbf{q} = -\kappa_c \nabla T$. 
A temperature gradient is
imposed along each C$\rightarrow$P cell by using a set of Nos\'e-Hoover
thermostats (NHT) placed along the direction of the heat flux. The
outer and inner thermostats are located a distance $d$ apart, and the
temperature difference between both set to $d\, \nabla T \cdot \mathbf {n}$.  
Typically, at each C$\rightarrow$P cell we have used a set of two
or three NHT's along a distance of $3\sigma$ or $4\sigma$.  The values of the
$Q$-parameter appearing in the NHT formulation \cite{Frenkel.book} 
were chosen small enough to minimise unphysical dynamics,
i.e., we have chosen $Q\simeq 5$. The main benefit of using the Nos\'e-Hoover
formulation is the small distortion these thermostats 
introduce to the particle dynamics compared with 
other ways of implementing thermostatting \cite{Frenkel.book}.

\subsection{Mass exchange: particle insertion}
One important condition on the particle insertion, inherited from the
balance of potential energy, is $\left<\psi^{\prime}\right>=\phi$ (see
Sec. \ref{enerbal}).  To deal with this task, our strategy has been to
place the new particles in positions where $\psi^{\prime} \simeq
\phi$. As $\phi$ is, roughly speaking, the energy needed to insert a
new particle, we shall insert particles with energies close to the
local chemical potential.  In any case, this implies that the
insertions need to be made in very precise positions which depend on
the configuration of the rest of the particles.  To this end we have
developed an algorithm ({\sc usher}) that guides each new inserted
particle to a position where the potential energy is equal to $\phi$
(up to a pre-specified threshold). A brief explanation is given below
(see Delgado-Buscalioni and Coveney \cite{INSERT} for further
details).  We first note that while the {\sc usher} algorithm guides a
new particle to a correct location, the rest of the particles remain
frozen in position. {\sc usher} essentially performs the following
steps:
\begin{enumerate}
\item Place the new particle ($i=N+1$) at an initial position inside 
C$\rightarrow$P, $\mathbf{r}^{(0)}$.
\item Evaluate $\mathbf{f}_{N+1}= \sum_{j=1}^{N} \mathbf{f}_{N+1,j}$ 
and $$\delta t_{\sigma}= {\sqrt \frac {2 \delta r}{|\mathbf{f}_{N+1}|}}$$
Typically one can use $\delta r\simeq \sigma$. 
\item Move the new particle according to the update rule
$$ \mathbf{r}^{(n+1)}= \mathbf{r}^{(n)} + \frac{1}{2}  \mathbf{f}_{N+1}^{(n)} \delta t^2$$
where $\delta t= \min(\Delta_t,\delta t_{\sigma})$, with $\Delta_t \simeq 0.05$ in reduced units
\item Evaluate the relative difference between the 
specific internal energy of the new particle, $\psi^{\prime}_{N+1}$, and that prescribed by
the continuum, $\phi$: $Err=|\psi^{\prime}_{N+1}-\phi|/|\phi|$
\item The particle is correctly inserted if $Err$ is small enough (typically $\sim 0.1$).
\end{enumerate}

Let us show how the {\sc usher} algorithm easily overcomes the problem
of possible initial overlap with pre-existing particles. An overlap
leads to very large values of the interparticle force, $f_{N+1}>>1$,
so in this case $\delta t_{\sigma}<<1$ and $\delta t=\delta
t_{\sigma}$.  But by construction, during the interval $\delta
t_{\sigma}$ the new particle moves a distance of the order of the
particle size $\sigma$ in the direction of minimum energy, just enough
to avoid any initial overlap. Then, as the particle steadily moves
towards a local minimum of energy, $f_{N+1}$ decreases, and $\delta
t_{\sigma}$ increases until it becomes larger than $\Delta_t$. Then,
$\delta t=\Delta_t$ is fixed. 
For liquid densities varying between $\rho=0.5-0.8$, the {\sc usher}
algorithm typically needs 15-90 iterations (single-force evaluations)
to correctly place a new particle \cite{note}
. By introducing the particles with
$\psi^{\prime}=\phi(1 \pm Err)$, we found that, upon averaging over
$\Delta t_C$, the condition $\left<\psi^{\prime}\right>=\phi$ holds
within about $2\%$ (even using values of $Err$ as large as $0.5$).

Until now we have not mentioned any limitation on the sizes of the
coarse mesh and time step $\Delta X$ and $\delta t_C$. 
Comments on this topic are very scarce in the previous literature on
hybrid methods for fluids.  Moreover, since the local averages 
are made using these spatial and temporal windows,  it is
also appropriate to formulate any condition on $\Delta X$ and $\Delta t_C$
before presenting the results of the tests carried out for different
flows.

\section{\label{req} Length and time scale prerequisites}
\subsubsection{Arising from consideration of the microscopic description}
Continuum fluid dynamics rest on the local equilibrium assumption.  
This means that, in
order to define a local thermodynamic and hydrodynamic state
characterising the coarse-grained variables at 
the C$\rightarrow$P and P$\rightarrow$C cells, the
size of these cells needs to be greater than the mean free path
$\lambda$ extracted from the particle dynamics.  Moreover the local
equilibrium within each cell should be attained on times scales
smaller than $\Delta t_C$. This means that $\Delta t_C$  has to be larger than 
the collision time $\tau_{col}$. In summary, $\Delta X > \lambda$ and
$\Delta t_C> \tau_{col}$.  In the case of a LJ-fluid it is possible to use
the hard-sphere approximation to make an order-of-magnitude estimate,
$\Delta X > 0.2\rho^{-1}$ and $\Delta t > 0.14 \rho^{-1}T^{-1/2}$. 
These conditions become less restrictive at larger densities;
as an example, for $T=1$, $\rho=0.5$
and a typical MD time-step $\Delta t_P\sim 10^{-3}$,
local equilibrium would require
$\Delta t_C/\Delta t_P \geq 100$ integration steps.

\subsubsection{Arising from consideration of the hydrodynamic description
\label{reqh}}
Conditions on $\Delta X$ and $\Delta t_C$ are firstly imposed by the
smallest characteristic length and time scales involved in the process
under investigation (say $2\pi/k_{\max}$ and $2\pi/\omega_{\max}$
respectively).  Practically, to correctly recover the smallest
spatio-temporal flow pattern one needs at least 8 points per period,
so $\Delta t_C \leq \pi/4\omega_{\max}$ and $\Delta X \leq
\pi/4k_{\max}$.  The numerical stability of the C-solver algorithm may
also impose limitations.  As mentioned in Sec. I, algorithms with
explicit time discretisation are better suited for the C-solver of a
hybrid scheme.  A necessary condition for their numerical stability is
$\mathrm{C}=U \Delta t_C/\Delta X\leq 1/2$, where C is the Courant
number and $U$ the maximum characteristic flow velocity.  The value of
$U$ depends on the physical process one is dealing with but, to
provide numbers in the present discussion, let us assume that we are
dealing with low or moderate Reynolds numbers.  Then, if the process
is a diffusive one, $U=\nu/\Delta X$; alternatively, if sound waves
are relevant within the flow, $U=c_s$.  In summary, the computational
window for $\Delta t_C$ should be, $0.14\rho^{-1} T^{-1/2}<\Delta t_C
\leq \Delta X/(2U)$.  Using the maximum grid-spacing allowed, $\Delta
X=\pi/4k_{\max}$, one obtains the computational windows for $\Delta
t_C$ shown in Fig. 2 {\em versus} $\rho$, for
$k_{\max}=\left\{0.1,0.2\right\}$, $U=\left\{\nu/\Delta X,c_s\right\}$
and $T=2.5$.  As expected, a sound wave requires smaller time steps
than a diffusive process.  For large enough $k_{\max}$, the temporal
and spatial computational window may be highly localised; e.g. for
$\rho=0.5$ one should use $0.2<\Delta t_C < 0.5$ if waves with
wavelength smaller than $30\sigma$ need to be captured by the
coarse-grained description.  As the density decreases these conditions
become much more restrictive, until the acoustic time finally becomes
smaller than the collision time (see Fig. 2).  Also, in rarefied
gases, if $\rho<k_{max}/4$, the mean free path becomes larger than the
wavelength, but here we shall not be concerned with situations where
the Navier-Stokes equations are not appropriate (see Garcia {\em et
al.}  for further discussion \cite{Gar00}).

\section{\label{hyd}
Tests: Hydrodynamic modes}

As already mentioned our hybrid scheme has been tested under stationary and
unsteady flows. Typical stationary nonequilibrium states were considered,
such as heat conduction profiles \cite{Ciccotti} and 
Couette profiles \cite{Flek00,Ciccotti}.
The microscopic reconstruction of these flows has been well 
studied in the literature that nothing new may added here. In passing, we note
the transient times
to achieve the steady state from the rest solution were found to be in agreement
with the diffusive times $L_x/\kappa$ and $L_x/\nu$.
The rest of the discussion will be focused on our choice of unsteady scenarios,
which are described by the decay of transversal and longitudinal waves.
These flows are now briefly presented, using standard hydrodynamics.

Consider a fluid at equilibrium characterised by homogeneous mass
density, $\rho^e$, specific energy $e^e$ and a vanishing mean velocity
$\mathbf{u}^e=0$.  Our procedure is to perturb this equilibrium state
with different hydrodynamic fields $\left\{\rho^p,\mathbf{u},e^p
\right\}$ (periodic in the $x$ direction, i.e. $\mathbf{k}=k \mathbf{i}$) and then make use of the C$\rightarrow$P coupling scheme described in Sec. \ref{3} to verify that,
within the  particle region, the subsequent evolution
towards equilibrium is carried out in a hydrodynamically consistent  way.
If use is to be made of the linear hydrodynamic theory, the externally induced
perturbations should be small enough to guarantee that the relaxation process is
always governed by the linearised mass, momentum and energy equations
(\ref{mass})-(\ref{ener}) \cite{HanMcD}. We defer further discussion of this point to
Sec. \ref{4} below.  

As is customary, to solve the linearised set of equations a Laplace-Fourier 
transform (LFT) is first performed \cite{HanMcD}.  The LFT of any perturbative variable, say
$\Phi(\mathbf{r},t)$, will be denoted as
\begin{eqnarray}
\Phi(\mathbf{k},t)&\equiv& \int_{-\infty}^{\infty} \Phi(\mathbf{r},t) \exp(-i\mathbf{k\cdot r}) d\mathbf{r},\\
\widehat{\Phi}(\mathbf{k},z)&\equiv&\int_0^{\infty} dz \exp(izt) \Phi(\mathbf{k},t) dt.
\end{eqnarray}
The LFT of the linearised equation (\ref{mass})-(\ref{ener}) 
leads to the following algebraic system 
for  $\widehat{\mathbf{\Phi}}=\left(\widehat{\rho}^p,\widehat{T}^p,\widehat{j_x^p},\widehat{j_y^p},\widehat{j_z^p}\right)$ \cite{HanMcD},
\begin{equation}
\label{pertu.eq}
\mathbf{M} \widehat{\mathbf{\Phi}}^T (\mathbf{k},z) =\mathbf{\Phi}^T(\mathbf{k},0),
\end{equation}
where the hydrodynamic matrix is,
\begin{eqnarray}
\mathbf{M}&=&
\left(
\begin{array}{ccccc}
 -i z &  0 & ik & 0 & 0\\
0 & -iz+ \kappa \gamma k^2 & ik \frac{\gamma -1}{\rho^e\alpha} & 0 & 0\\
ik c_s^2/\gamma & ikD & -iz + b k^2 & 0 & 0\\
0&  0 & 0 & -iz +\nu k^2 &0 \\
0 & 0 &0 & 0 & -iz +\nu k^2 
\end{array}
\right).
\label{linearLF}
\end{eqnarray}
We note that instead of using $e^p$,
the energy equation is expressed in terms of temperature fluctuations.
Also, for clarity it is better to write  the solution $\bm{\widehat{\Phi}}$ in
terms of the $t=0$ Fourier-transformed perturbative heat  density  $Q^p$
and pressure $P^p$.  These quantities are related to $\rho^p$ and $T^p$ through
the relations
\begin{eqnarray}
\label{e(r,t)}
e^p&=&  c_v T^p + \left(\frac{\partial e}{\partial \rho}\right)_{T^e} \rho^p, \\
\label{Q(r,t)}
Q^p&=& \rho^e c_v\left(T^p - \frac{\gamma-1}{\rho^e \alpha} \rho^p \right),\\
\label{P(r,t)}
P^p&=& \frac{m c_s^2}{\gamma} \left(\rho^e \alpha T^p - \rho^p \right).
\end{eqnarray}
Here $c_p$ and $c_v$ are the specific heat at constant pressure and volume, respectively;
$\gamma=c_p/c_v$, is the adiabatic coefficient;
$\kappa=\kappa_c/c_p\rho^e$ is the thermal diffusivity; $\alpha=-(\partial \rho/\partial
T)_{P}/\rho^e$ the thermal expansion coefficient; and the kinematic longitudinal
viscosity, $b=(4\eta/3 +\xi)/\rho^e$ is related to the sound attenuation
coefficient, $\Gamma$, through $2 \Gamma=b +(\gamma-1) \kappa$. 
Finally $D=(\partial
P/\partial T)_{\rho}$ and the adiabatic speed of sound is $c_s^2=\gamma
\left(\partial P \partial\rho \right)_{T}$.

Provided that, in the hydrodynamic limit, the wavelengths of the perturbations
are much larger than the mean interparticle distance, it is sufficient to
obtain the solution of Eq. (\ref{pertu.eq}) up to $O(k^2)$ \cite{HanMcD}.
Putting $z \rightarrow \omega \in \cal{R} $ into Eqs. (\ref{pertu.eq}) and (\ref{linearLF}),
leads, after some algebra, to the following identities
\begin{eqnarray}
\label{LFT}
\widehat{T}(\mathbf{k},\omega)&=& \frac{E_{\kappa} Q(\mathbf{k},0)}{\rho^e c_p} +
\frac{\gamma-1}{\rho^e \alpha} E_{SR} P(\mathbf{k},0) + 
\frac{\gamma-1}{\rho^e \alpha c_s} E_{SI} j_x(\mathbf{k},0), \\ 
\label{LFrho}
\widehat{\rho}(\mathbf{k},\omega)&=& -\frac{\alpha}{c_p} E_{\kappa}  Q(\mathbf{k},0) + 
\frac{1}{m c_s^2} E_{SR}  P(\mathbf{k},0) + 
\frac{1}{c_s} E_{SI} j_x(\mathbf{k},0), \\ 
\label{LFjx}
\widehat{j}_x(\mathbf{k},\omega)&=& \frac{1}{m c_s} E_{SI} P(\mathbf{k},0) + 
E_{SR} j_{x}(\mathbf{k},0),\\
\label{LFjy}
\widehat{j}_y(\mathbf{k},\omega)&=& E_{\nu} j_y(\mathbf{k},0),\\
\label{LFjz}
\widehat{j}_z(\mathbf{k},\omega)&=& E_{\nu} j_z(\mathbf{k},0),
\end{eqnarray}
and, using Eqs. (\ref{Q(r,t)}) and (\ref{P(r,t)}),
\begin{eqnarray}
\label{LFQ}
\widehat{Q}(\mathbf{k},\omega)&=& E_{\kappa}Q(\mathbf{k},0), \\
\label{LFP}
\widehat{P}(\mathbf{k},\omega)&=& E_{SR} P(\mathbf{k},0) + 
mc_s E_{SI} j_x(\mathbf{k},0),
\end{eqnarray}
where the following propagators have been introduced,
\begin{eqnarray}
\label{ed}
E_{\kappa}(k,\omega)&=&\exp(-\kappa k^2 t),\\
\label{esr}
E_{SR}(k,\omega)&=&\exp(-\Gamma k^2 t) \cos(c_s k t),\\
\label{esi}
E_{SI}(k,\omega)&=& -i \exp(-\Gamma k^2 t) \sin(c_s k t),\\
\label{enu}
E_{\nu}(k,\omega)&=&\exp(-\nu k^2 t).
\end{eqnarray}

The shear or transverse modes correspond to momentum perturbations along either $y$
and/or $z$ axis (i.e.  perpendicular to the wavevector $k\mathbf{i}$); from
Eqs. (\ref{LFjy}) and (\ref{LFjz}) it is clear that they are completely decoupled.  The 
remaining hydrodynamic variables $\left\{\rho^p,T^p,j_x^p\right\}$ are 
coupled and conform to three longitudinal 
modes which can be divided into two subgroups: two sound modes
(involving longitudinal momentum and pressure at constant entropy) and one heat mode
(involving heat diffusion).  The fact that the heat density
is an independent mode is evident from
Eq. (\ref{LFQ}), where it is seen that it relaxes diffusively
$\propto \exp(-\kappa k^2 t)$.  We note that $Q^p=T^e s^p$ is
essentially the fluctuation of the entropy density $s^p$ \cite{HanMcD}.
From Eqs. (\ref{LFjx}) and (\ref{LFP}) it is easily shown that the two sound modes
($\widehat{P}/c_s \pm  \widehat{j}_x$) decay like $\exp(ik c_s t -\Gamma k^2t)$.

\section{\label{4} Results}

\subsection{Set-up and initial states}
The coupling scheme was implemented and tested in the 
set-up shown in Fig. 3. The system is periodic
along $y$ and $z$ directions and 
the gradients of the continuum variables are set along the $x$
direction. The particle region occupies a 
region of size $L_x$ around $x=0$ and of 
size $L_y=L_z$ along the periodic directions. The
P region is divided into control cells of size $\Delta X$
where in local averages are taken.  The centres of the two C$\rightarrow$P slabs
(the outermost cells) are situated at $x=\pm |L_x-\Delta X/2|$.
The deviation from the local equilibrium assumption was monitored in terms of
the relative difference of the cell-averaged pressure and energy
with respect the values given by the equation of state of
Johnson {\em et al.}  Ref \cite{EOS}.
Around a distance $1.5\sigma$ away from the C$\rightarrow$P interface the
typical maximum deviations were only about $6\%$.

The initial perturbative flow was prepared by first letting the P region
relax until a vanishing and homogeneous mean flow was obtained. Then,
during a small time interval ($\sim 3\tau$), the particle velocities
were periodically changed according to a Maxwellian distribution with the
desired velocity profile and local cell temperature. The
resulting initial state was then analysed to extract the Fourier
components of the whole set of flow variables ($\mathbf{v}$, $\rho$,
$T$, $e$, P). For the sake of consistency these were extracted by
Fourier transforms of the cell-averaged variables 
\begin{eqnarray}
\label{fourier}
\bar{\phi}^{(n)}_{\cos}(t) \equiv\frac{c_n}{M_c}\sum_{l}^{M_c} \bar{\phi}(X_{l},t) \cos(k_n X_l), \\
\label{fourier.s}
\bar{\phi}^{(n)}_{\sin}(t) \equiv\frac{c_n}{M_c}\sum_{l}^{M_c} \bar{\phi}(X_{l},t) \sin(k_n X_l),
\end{eqnarray}
where $k_n=n\,k$ ($n\in \mathcal{N})$ and; $c_n=1$ for $n=0$, $c_n=2$ otherwise.
In any case, it was checked 
that the Fourier transform of the 
microscopic variables $\phi^{(n)}=c_n\sum_{i}^{N} \phi(x_i,t) \exp(-ik_n x_i)/N$
yields essentially the same output as Eqs. (\ref{fourier}) and (\ref{fourier.s}).

The initial Fourier transforms calculated from Eqs. (\ref{fourier})
were injected into Eqs. (\ref{LFT})-(\ref{LFP}) to obtain the
time-evolution of the continuum variables.  These, in turn, were used
to calculate the fluxes imposed on the C$\rightarrow$P cells over
time.  The transport coefficients used were those reported in the
literature \cite{TRANSP,EOS}.  As an interesting check, it was found
that the coupling scheme failed significantly if the transport
coefficients used in the C-region differed by more than about $15\%$
from those of the LJ fluid.  In particular, the oscillation frequency
of the averaged particle velocity differed with respect to that of C,
while correlations decayed at a faster rate than those of the C-flow.
Therefore, in cases where the constitutive relations are not known,
this result means that it is first necessary to measure the transport
coefficients from the particle dynamics (using any standard molecular
technique), before applying the hybrid scheme, particularly if
unsteady flows are to be studied.

The wavelengths of the initial perturbations were chosen to be much
larger than the mean free path, i.e. $2\pi \lambda/k <<1$, in order to
work within the hydrodynamic regime. In other words, the dependence of
the transport coefficients on the wavenumber was negligible
\cite{HanMcD}. The amplitudes of the initial perturbation were chosen
small enough to ensure that the subsequent relaxation process could be
described by linear theory. In particular, if $\bar{\mathbf
v}^{(1)}(t)$ is the maximum Fourier amplitude of the velocity, the
typical values of the Reynolds number at $t=0$ [Re$=|\bar{\mathbf
v}^{(1)}|\rho_e/(k\eta)$] were Re$(0)\leq 3$. As $|\bar{\mathbf
v}^{(1)}(t)|$ decays exponentially, convection was present only in the
first stages of the relaxing flow, but it was not strong enough to
produce significant deviations from the linear theory (non-linear
effects become dominant for Re$>O(10)$ \cite{TRITTON}).  The maximum
Mach number was less than $0.2$, and density fluctuations were around
$|\bar {\rho}^{(1)}|/\rho_e \sim \mathrm{Ma}^2\simeq 0.05$.


In another test the hybrid scheme was applied to a fluid in mechanical
($\mathbf{u}=0$) and thermodynamical equilibrium ($\rho=0.5$, $T=3.5$,
$e= 2.7$, $P=3.2$) during a longer simulation (50$\tau$) to check for
any possible spurious drift in the overall momentum and energy 
(note that Eq. (\ref{flux.mass}) ensures the mass conservation by
construction). During this calculation, the total momentum inside the
P region was conserved up to $5\times 10^{-4}$ and the total energy
fluctuated $\sim 5\%$ around its equilibrium value.  The size of
these fluctuations is consistent with the system size (which
contained $N=1600$ particles and a specific heat of $c_v=1.8$).  Note
that the total energy of the system cannot be conserved because a part the
system is connected to a thermostat and it also receives mechanical
energy from C.


\subsection{Transversal waves}

In order to test the transfer of momentum flux along the direction
perpendicular to the C$\rightarrow$P interface, planar shear waves along $x$
were excited in a LJ-fluid with $\rho=0.5$, $T=2.5$ and $\eta=0.75
\pm 0.05$ (the error bar comes from Ref.\cite{TRANSP}).  These waves were created by
imposing sinusoidal $y$-velocity profiles $v_y(x)=v_{y,\sin}^{(1)}\sin(k x)$.  

Figure 4 shows results for perturbations with $v_{y,\sin}^{(1)}=1.0$
in a system with $L_x=20\sigma$, $L_y=L_z=7\sigma$ and $M_c=10$ cells
($\Delta X=2$).  The filled circles in Figure 4a correspond to the
main Fourier component of the velocity $\widehat{v}_{y,\sin}^{(1)}$
for a calculation using wavenumber $k=0.31$.  The relaxation process
is indeed exponential, until the noise-to-signal ratio becomes large
enough (around $t>10$) and, the observed decay rate $(0.14\pm
0.1)\tau^{-1}$ agrees perfectly with the theoretical value
$k^2\eta/\rho^{e}$.  To explicitly appreciate the effect of the hybrid
coupling we show (open circles in Fig. 4a) the outcome of a
calculation in which the unsteady shear stress contribution to the
external force $F^{ext}$ was set to zero, leaving just the
contribution of the equilibrium hydrostatic pressure.  As expected, if
less momentum flux is provided towards P, the decay rate is
appreciably slower ($0.07\tau^{-1}$) than the hydrodynamic one
($0.14\tau^{-1}$).  Figure 4b shows the autocorrelation function (ACF)
of $\widehat{v}_{y,\sin}^{(1)}(t)$ for another perturbation with an
slightly different wavenumber, $k=0.35$.  The good agreement with the
theoretical decay [$\exp(-0.17t)$, in dashed line] shows that the
scheme is able to deal with small variations in the perturbation
shape.

\subsection{Longitudinal perturbations}

Longitudinal waves transport mass, momentum and both mechanical and
thermal energy; they are therefore perfectly suited for an overall
test of the hybrid scheme behaviour.  In our set-up the wavevector of
these waves was perpendicular to the C$\rightarrow$P interface and
they were generated by imposing mean velocities along the $x$
direction as either pure cosinusoidal, sinusoidal profiles, or
combinations of both profiles. As explained before, the particle
velocities were extracted from Maxwell-Boltzmann distributions at the
local mean velocities and at a constant temperature.  The mean
velocity profile induces pressure, density and energy fluctuations
with a periodic pattern in the $x$ direction.  The temporal behaviour
of the main flow variables is now described.

\subsubsection{Mass}

One of the main problems involved in the hybrid mass transfer at
C$\rightarrow$P is that although the continuum flux is a floating point number,
one can only possibly exchange an integer number of  particles.
In order to adhere closely as possible to the prescribed  continuum
mass flux, the following procedure is followed during
each interval $t_C< t\leq t_C+\Delta t_C$, $t_C=m\Delta t_C$, $m\in \mathcal{N}$.
First one evaluates the quantity
\begin{equation}
\xi(t_C)\equiv \int_{t_C}^{t_C+\Delta t_C} s(t) dt.
\end{equation}
This floating point number, which represents the number of particles that should cross 
the C$\rightarrow$P interface along $t_C< t\leq t_C+\Delta t_C$,
is converted into an integer $\delta N(t)$ by the following construction:
\begin{eqnarray}
\delta N(t) &=& \mathrm{NINT}\left[\xi(t_C)\right] + \delta{\xi}(t-t_k) \\
&&\delta{\xi}(t-t_k) =\mathrm{INT}\left[\int_{t_k}^{t-\Delta t_P} (\xi(t^{\prime}) -\delta N(t^{\prime}))\,dt^{\prime}\right]
\end{eqnarray}
where $t_k$ is such that $|\delta{\xi}(t-t_k)|\leq 1$ and
$0=t_0<t_k<t_{k+1}$. The deviation $\delta{\xi}(t-t_k)$ assimilates
the errors made through successive rounding off ($\xi\rightarrow
\mathrm{NINT}[\xi]$). When $|\delta{\xi}(t-t_k)|$ becomes larger than
one (at $t=\left\{t_k\right\}$) a particle is added to (or extracted
from) $\delta N$ and the corrector $\delta{\xi}$ is then reset to
zero. To minimise the effect on the remaining particles over each
interval $\Delta t_C$, the particle crossings are regularly separated
in time, at a rate as close as possible to $\delta N(t_C)/\Delta t_C$.
As illustrated in Fig. 5, this kind of procedure enables us to follow
rather closely the desired mass flux.

\subsubsection{\label{momt} Momentum}
Figures 7, 8 and 9 show the time dependence of the Fourier components
of the main hydrodynamic and thermodynamic variables.  Dashed lines
correspond to the theoretical trends obtained via the inverse Fourier
transform of Eqs. (\ref{LFT})-(\ref{LFP}).

For the reasons explained in Sec \ref{uns}, it is particularly
important to ascertain that the averaged velocity at C$\rightarrow$P
correctly follows the desired continuum flow velocity. Figure 6 shows
the instantaneous $\bar{v_x}$ and time averaged velocity $\left<
\bar{v_x}\right>$ at both C$\rightarrow$P cells.  We obtained very
good agreement, the deviation of $\left<\bar{v}_x\right>$ with respect
the continuum value being less than about $5\%$ along most part of the
damped oscillation (see Fig. 6).  The importance of evaluating the
continuum fluxes at precisely the C$\rightarrow$P interface ($x_W$ in
Fig. 1), is illustrated in Fig. 6 by comparison with the outcome of a
calculation in which the fluxes are evaluated at $x=x_0$. For this
calculation, the deviation with respect to the continuum prescription
is clear and its magnitude agrees with the estimate made in
Sec. \ref{uns}, $\sim 30\%$.

Figure 7a  shows the time evolution  of the Fourier  amplitudes of the
velocity in one of  the calculations (a pure cosinusoidal perturbation
with    $\bar{v}_{x,\cos}^{(1)}=0.6$).      The    overall    velocity
$\bar{v}_{x,\cos}^{(0)}$  remains close to  zero, confirming  that the
method     conserves    the     initial    total     momentum.     The
$\bar{v}_{x,\sin}^{(1)}$  component (initially set  to zero)  has been
also  included to  display the  level of  noise.   The autocorrelation
function    (ACF)    of    $\bar{v}_x^{(1)}=\bar{v}_{(x,\cos)}^{(1)}+i
\bar{v}_{(x,\sin)}^{(1)}$  is  shown in  Fig.  7b  for  two runs  with
different  initial  profiles (for  more  details  see  the caption  of
Fig.  7).   These data  were  fitted  to  the hydrodynamic  expression
arising from Eq. (\ref{LFjx}) and in  Fig. 7 these fits are shown with
dashed lines.  The best fit  to the velocity was obtained with $\Gamma
k^2 = 0.076$ and $c_s k  =0.867$, while for the ACF it yielded $\Gamma
k^2 =0.072$  and $c_s  k =0.867$.  These  values coincide,  within the
error bars, with those imposed  by the C-flow (obtained upon insertion
of   the   transport   coefficients   reported   in   the   literature
\cite{TRANSP}) $0.071$ and $0.88$, respectively.

\subsubsection{Thermodynamic variables}

We start by observing that, unlike longitudinal momentum and pressure,
the relaxation of density, temperature and energy perturbations
includes not only an acoustic part, but also an entropic contribution
proportional to $\exp(-\kappa k^2 t)$ [see
Eqs. (\ref{LFT})-(\ref{enu})].  However, as the initial perturbation
considered was a mechanical one ($T^{(n)}\simeq 0$), the entropic
contribution is rather small.  This can be seen in the theoretical
expressions written in Fig. 6: the amplitude at $t=0$ of the entropic
part of any heat-related variable is nearly six times smaller than its
mechanical counterpart.  This observation led us to a more careful
study of the effect of heat conduction.  As explained in Sec
\ref{cond}, heat currents through each C$\rightarrow$P cell were
created by imposing several (typically two) Nos\'e-Hoover thermostats
(NHT's) placed close to the C$\rightarrow$P interface over a distance
of $4\sigma$.  We found it very informative to study the effect of the
number $p$ of NHT's per C$\rightarrow$P cell (notation:
$p$-NHT$_{CP}$) on the collective behaviour of the system. In some
calculations, this number was reduced to merely 1-NHT$_{CP}$, in such
a way that only the local temperature prescribed by the continuum was
imposed (and not the heat flux).  The results of this comparison may
be seen in Figs. 8 and 9.  Calculations using 1-NHT$_{CP}$ yielded
essentially the same evolution of the velocity and pressure as those
with a larger number of NHT's, although the pressure for the
1-NHT$_{CP}$ case showed a slight phase-lag, see Fig. 8d.  This is not
surprising as $\mathbf{v}$ and P are governed by acoustic terms and
are independent of heat transfers. Using 1-NHT$_{CP}$, deviations on
$\rho^p$, $e^p$ and $T^p$ with respect to the hydrodynamic trend are
indeed appreciable, while results with 2-NHT$_{CP}$ adhere closely to
the analytical curves (see Fig. 8).  In any case, the calculation of
the entropy production is the best way to highlight the completely
different qualitative behaviour of 1-NHT$_{CP}$ with respect to two
(or more) NHT$_{CP}$.

\subsubsection{Entropy}	

The entropy fluctuations, or more precisely the perturbation of heat $Q^p=T^e s^p$, was
calculated using Eq. (\ref{Q(r,t)}).  We note that, in order to reproduce the steady
diffusive heat decay in Eq. (\ref{LFrho}), an exact cancellation of the acoustic
oscillations coming from $\rho^p$ and $T^p$ is necessary [see Eqs. (\ref{LFrho})and
(\ref{LFT})].  As stated above, the contribution associated with heat diffusion in
Eqs. (\ref{LFrho}) and (\ref{LFT}) is rather small compared to the mechanical one. This
observation indicates that small mechanically-driven fluctuations around the local
thermodynamic equilibrium may become large enough to alter the purely exponential heat
decay.  In other words, Eqs. (\ref{Q(r,t)}) and (\ref{LFQ}) provide a demanding test of the
coupling scheme under the present flow.  Figure 9 shows the main Fourier component of the
heat perturbation $Q^p$ obtained for 1- and 2-NHT$_{CP}$. The dashed lines correspond to
the theoretical expectation. It is evident that the 1-NHT$_{CP}$ case does not obey the
second law of thermodynamics at all. On the contrary, a rather good agreement with the
theoretical trend is obtained when using at least 2-NHT$_{CP}$.  In Fig. 9, it is seen that
the behaviour measured in the 2-NHT$_{CP}$ case exhibits fluctuations around the
theoretical straight line. Typically, the largest excursions last around $3 \tau$
from pure exponential decay. As previously stated, they may be due to the
weakness of the entropy perturbation, but to confirm this statement we plan 
in the future to study some ``heat-driven'' flows 
(such as a heat pulse with initial zero mean velocity).

\section{ \label{con}Conclusions}

We have presented the core of a hybrid continuum-particle method for 
moderate-to-large fluids which 
couples mass, momentum and energy transfers 
between two regions, C and P, 
described respectively by continuum fluid dynamics
and by discrete particle Newtonian dynamics.
Both domains overlap within a coupling region divided into two sub-cells 
which account for the two-way exchange: C$\rightarrow$P 
and P$\rightarrow$C. While the procedure at the P$\rightarrow$C cell is simply
to average the particle-based (mass, momentum and energy) fluxes 
in order to supply open boundary 
conditions to the C domain, the operations at
the C$\rightarrow$P  cell are much less straightforward as they need to
reconstruct a large number of (particles') degrees of freedom 
only from the knowledge  of  the  
three fluxes of conserved-quantities arising within C.
The present work has been concerned with extending the C$\rightarrow$P 
coupling to arbitrary rates of mass, momentum and energy transfer.
To this end, the proposed method have been tested under
unsteady flows which demand conformance 
to the whole set of conserved variable densities. In particular, we 
have considered the set of relaxing flows arising from hydrodynamics,
namely longitudinal and transversal waves.
We have followed the idea  proposed by Flekkoy {\em et al.} \cite{Flek00}, 
in the sense that 
the scheme is explicitly based on direct flux exchange between the C and P regions. 
In order to deal with unsteady scenarios, we have shown  that
the fluxes injected into the particle region from the continuum region
need to be measured exactly at the C$\rightarrow$P interface and 
not at the nodes of the continuum lattice.

The implementation of flux exchanges requires the supply of energy
currents to the particle system arising from the C domain due to
advection, dissipation and conduction.  To inject the correct amount
of advected energy, the particle-averaged specific energy at the
C$\rightarrow$P cell needs to be equal to the continuum value. This
can only be achieved if the new inserted particles are placed at
positions where the (inter-particle) potential energy equals the
C-specified internal energy per unit mass.  This severe condition has
been implemented by the {\sc usher} algorithm, whose purpose is
two-fold: to provide the correct mass transfer rate and to ensure the
balance of energy advection.  In the proposed scheme, the balance of
energy dissipation arises naturally, provided that the cell-averaged
velocity and the injected momentum flux equal their continuum
counterparts. This is made possible by applying the external force
according to a flat distribution, instead of a biased one as used by
Flekkoy {\em et al.}  \cite{Flek00} but as a result, the new particles
have to be inserted within a non-vanishing density environment. This
is sorted out by the {\sc usher} in a very efficient way.  Energy
conduction has been implemented by using a set of Nos\'e-Hoover
thermostats adjacent to the C$\rightarrow$P interface, whose
temperature and position are determined through the continuum local
temperature gradient.  Confirmation of the validity of this procedure
is obtained from the correct rate of entropy production computed in
our simulations of longitudinal waves.  We showed that using only one
thermostat per C$\rightarrow$P cell (i.e. providing only the local
value of $T$ but not the heat flux) leads to negative entropy
production. Therefore, in the context of energy transfer, this result
reinforces the central importance of coupling-through-fluxes proposed
by Flekkoy {\em et al} \cite{Flek00}.

Enhancements to the present scheme are under investigation and merit
some discussion here. The number density at C$\rightarrow$P may be
controlled by a feedback algorithm which preserves the momentum flux
balance. Other kinds of implementations for the energy transport by
conduction also deserve to be considered.  Finally, we plan to
implement the P$\rightarrow$C coupling in conjunction with a finite
volume CFD solver in 3D. In order to extend the coupling scheme to
higher dimensions some additional complications will need to be
adressed.  First, a mass flux will be assigned to each cell within an
array of neighbourings C$\rightarrow$P cells.  To adhere to mass
continuity, particles will then have to be inserted within precisely
defined finite regions and the insertion algorithm may have to pay an
extra computational cost for this restriction of the search domain.  We have
checked, however, that the distance travelled by the {\sc usher}
algorithm from the initial trial position to the final insertion site
is rather small \cite{INSERT} (typically less than 1$\sigma$ and less
than 0.5$\sigma$ on average) so we do not expect any significant extra
cost if the search for insertion sites is done within volumes larger
than $(2\sigma)^3$.  In higher dimensions one may also have to smooth
to some extent the variations of the mean mechanical quantities
imposed along the C$\rightarrow$P region.  To this end, it may be
necessary to interpolate the external force along neighbouring
C$\rightarrow$P cells in such a way that the local momentum flux
imposed at each C$\rightarrow$P cell is still preserved.  In the same
way, although the Maxwell distribution, used here to choose the
velocities of the new particles, proved to be sufficient to ensure
momentum continuity for 1D coupling (i.e, with no neighbouring
C$\rightarrow$P cells), it may be convenient to use a Chapman-Enskog
distribution in higher dimensions.  This latter distribution enables
the average velocity of the inserted particles to conform to the
velocity gradient along neighbouring C$\rightarrow$P cells.  We hope
to report our findings in these areas in future publications.

\section{Acknowledgements}

We gratefully acknowledge fruitful discussions with E. Flekkoy,
P. Espa\~nol, G. Ciccotti and R. Winkler and useful comments from
B. Boghosian, A. Ladd and I. Paganobarraga. This work is supported by
the European Union through a Marie Curie Fellowship (HPMF-CT-2001-01210)
to RD-B.

\newpage

\begin{figure}[h]
\includegraphics{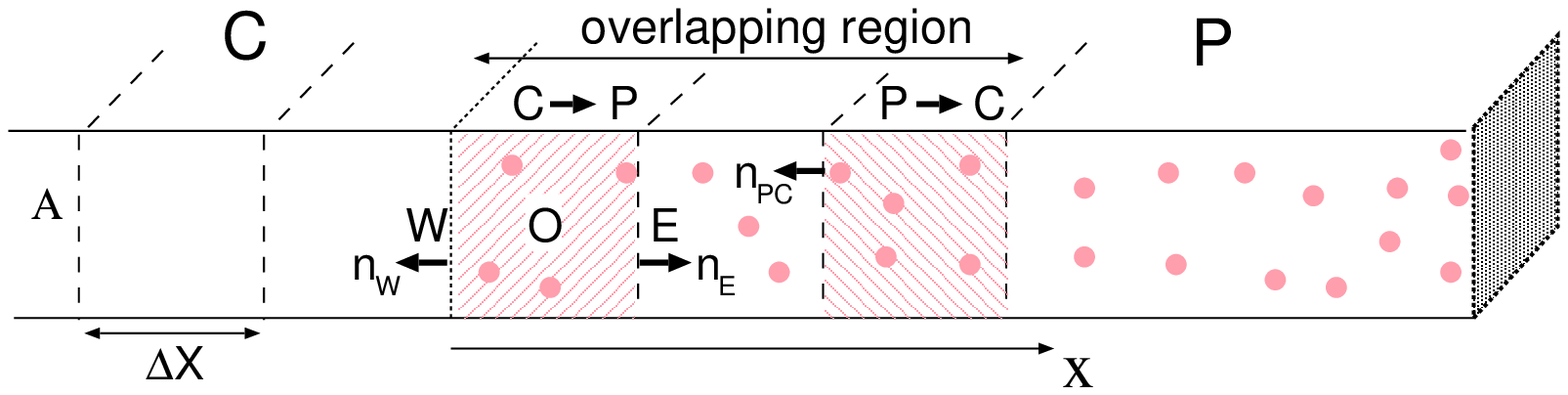}
\caption{(a) Spatial decomposition in our hybrid scheme. In this
example the P region is adjacent to a physical surface represented by
the rightmost shaded area.  The continuum region spans the space to
the left, at some distance from the surface.  The overlapping region
consists of a C$\rightarrow$P cell, where the C-flow is communicated
to P, and a P$\rightarrow$C cell, where particle averaged-fluxes are
injected into the C-flow. Dashed lines delimit the control cells of
the C solver, with area $A$ and grid-spacing $\Delta X$.  The letters
O, W and E denote the centre of a cell and its west and east surfaces,
respectively.  The main cell's vectors ($\mathbf{n}_W$, $\mathbf{n}_E$
and $\mathbf{n}_{PC}$) have been indicated (see text).}
\end{figure}

\begin{figure}[h]
\includegraphics{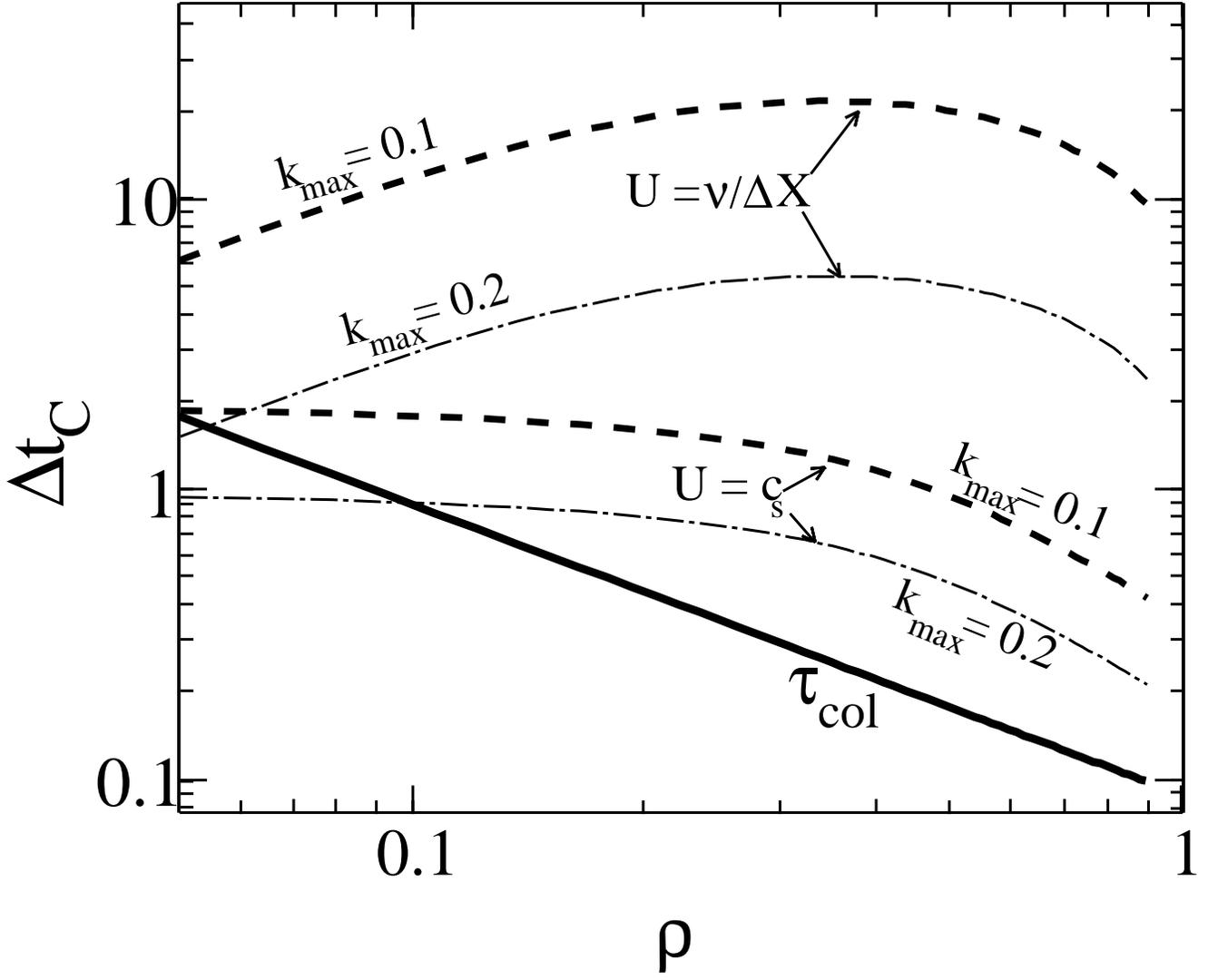}
\caption{Conditions imposed on the time step of the continuum solver
$\Delta t_C$ plotted {\em versus} the number density $\rho$. Variables
are expressed in the LJ reduced units ($\sigma$ for length and
$\tau=(\sigma^2 m/\epsilon)^{1/2}$ for time).  As discussed in
Sec. \ref{reqh}, $\Delta t_C$ has to be greater than the collision
time $\tau_{col}$ (the thick solid line) and smaller than $\Delta
X/(2U)$ (indicated with dashed and dash-dotted lines).  The typical
flow velocity $U$ is chosen to be either the sound velocity $U=c_s$ or
the diffusive velocity $U=\nu/L$, according to each case discussed in
Sec. \ref{reqh}.  The grid-spacing is $\Delta X=\pi/(4 k_{\max})$,
where $k_{\max}$ is the largest wavenumber to be captured within the
flow.
}
\end{figure}

\begin{figure}[h]
\includegraphics{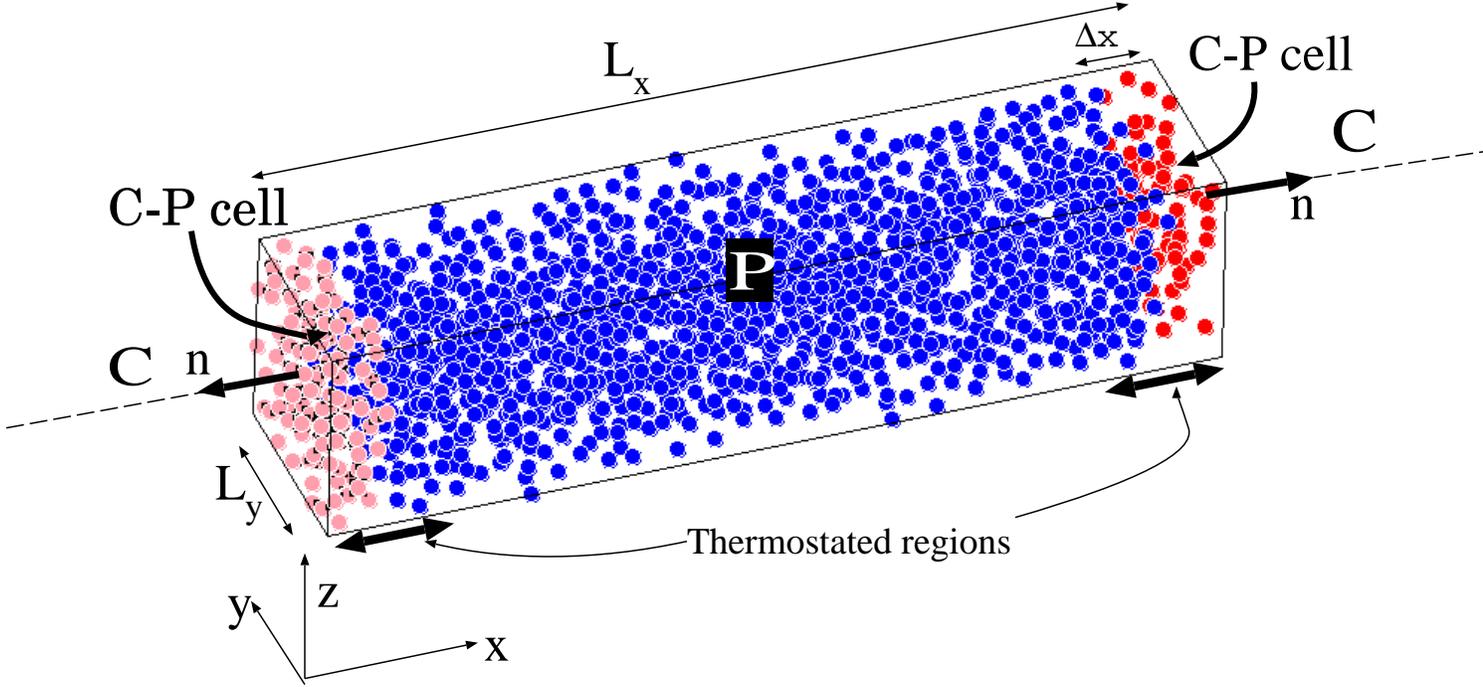}
\caption{The set-up for which the scheme has been tested (here
region P is surrounded by C). The fluxes of continuum variables are
imposed along the $x$ direction, while $y$ and $z$ are periodic.  The
hybrid coupling is applied at the C$\rightarrow$P cells and the heat
current is established along the Nos\'e-Hoover thermostatted regions
whose thickness is set between $(3-4)\sigma$.  We used $\Delta
X=(1-2)\sigma$, $L_x=(10-40)\sigma$ and $L_y=L_z=(7-9)\sigma$.
}
\end{figure}

\begin{figure}[h]
\includegraphics{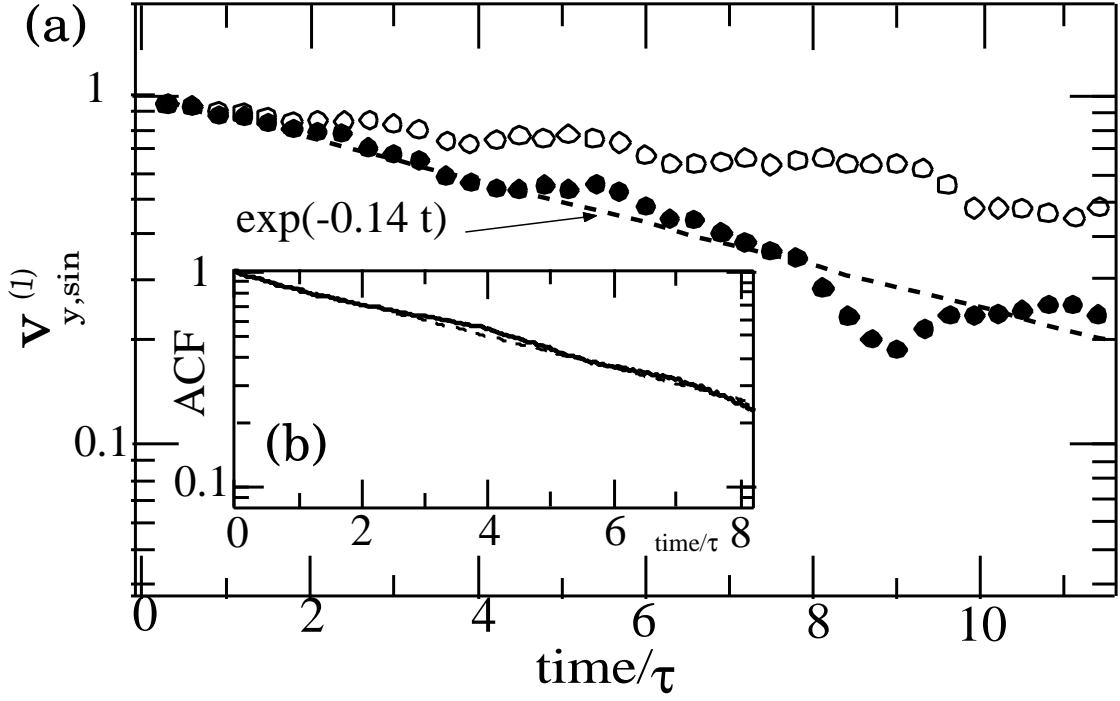}
\caption{(a) The main Fourier component of the cell-averaged
velocity $v^{(1)}_{y,\sin}$ (in units of $\sigma/\tau$, with $\tau=(\sigma^2
m/\epsilon)^{1/2}$).  
Results correspond to a transversal wave with wavenumber
$k=0.31$.
Comparison is made between a calculation in
which the full expression of the external force $F^{ext}$ was imposed
(filled circles) and another which did not include the viscous
contribution (open circles). The dashed line is the correct
hydrodynamic decay.  (b) The nondimensional autocorrelation function
(ACF) of $\bar{v}_{y,\sin}^{(1)}(t)$ for another transversal
perturbation with $k=0.35$, showing the theoretical decay in the
(partially hidden) dashed line. In all cases the initial amplitude was
$\bar{v}_{y,\sin}^{(1)}=1.0$ and $L_x=20\sigma$,
$L_y=L_z=7\sigma$. In abscissas, time is nondimensionalized with
the LJ reduced time unit $\tau=(\sigma^2 m/\epsilon)^{1/2}$.
}
\end{figure}

\begin{figure}[h]
\includegraphics{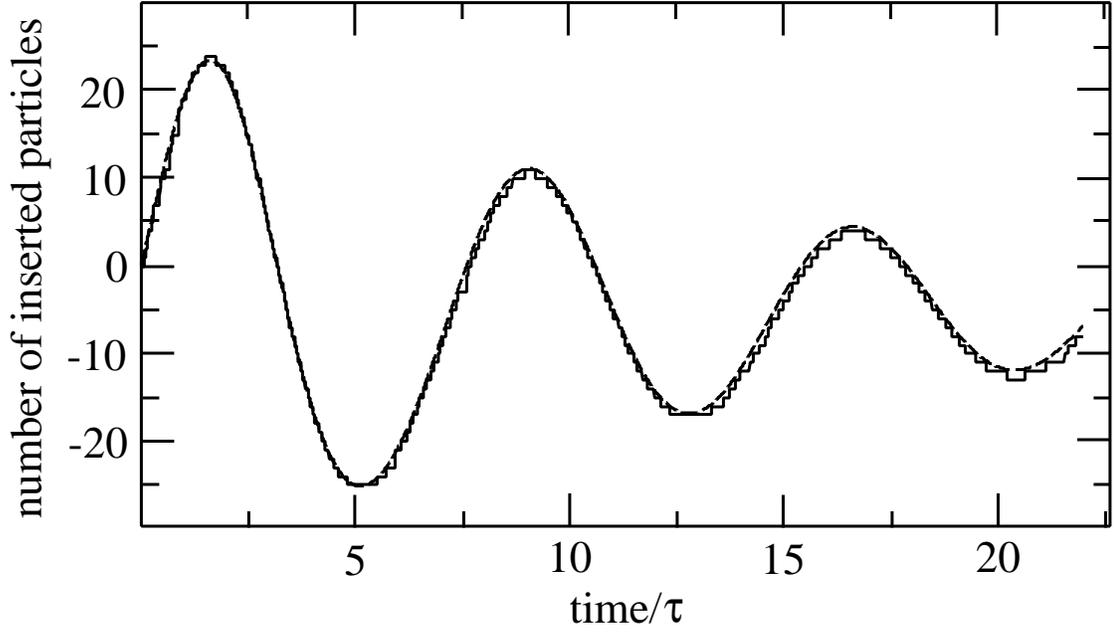}
\caption{The total number of inserted particles $\int_0^t \delta
N(t^{\prime}) dt^{\prime}$ at the rightmost C$\rightarrow$P cell,
along a simulation of a longitudinal wave with $k=0.168$ inside a
region ($L_x=40\sigma$, $L_y=L_z=9\sigma$) with $\rho^e=0.53$ and
$T^e=3.5$. The initial perturbative velocity profile was
$u_x=0.60\,\cos(kx)$.  The dashed line is the continuum prescription
$\int_0^t s(t^{\prime}) dt^{\prime}$ (see text). Time has been 
nondimensionalized with $\tau=(\sigma^2 m/\epsilon)^{1/2}$.
}
\end{figure}

\begin{figure}[h]
\includegraphics{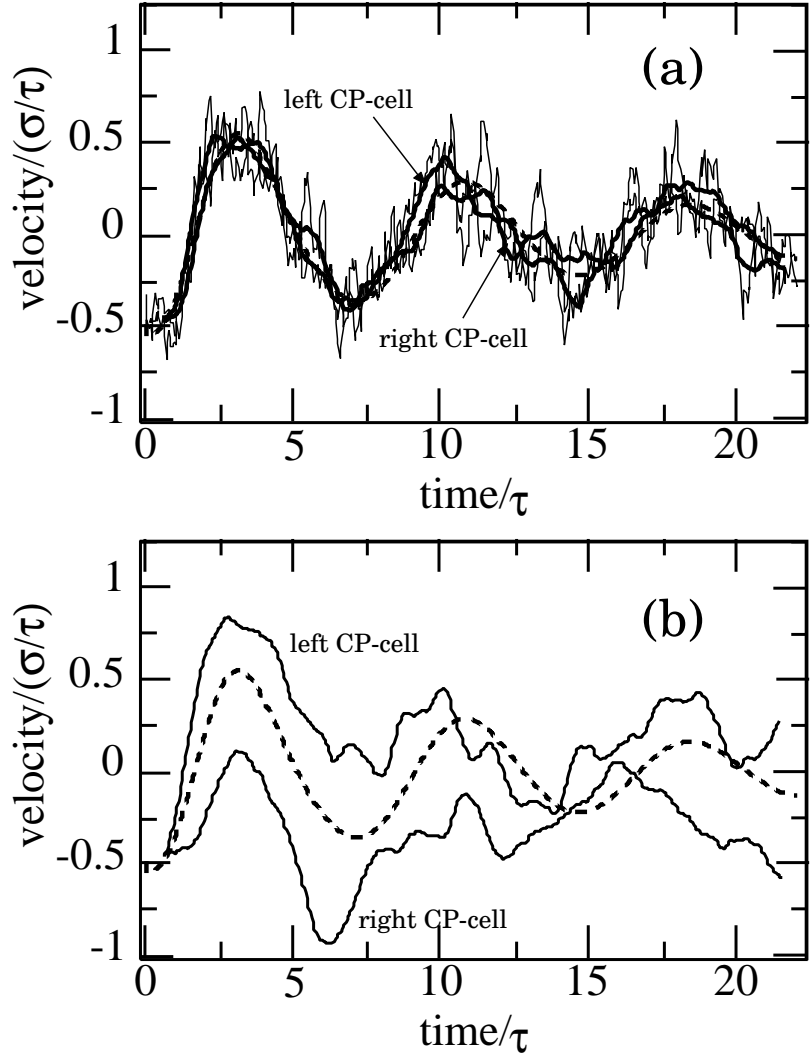}
\caption{The velocity at both C$\rightarrow$P cells for the same
parameters as in Fig. 5. The dashed line is the continuum prescription
(partially hidden).  In (a) we show the outcome of a calculation in
which the momentum flux from the C-region was evaluated at $x=x_W$
(see Fig. 1); the instantaneous $\bar{v}_x$ velocities are shown in
lighter dotted lines, while thicker solid lines are the time-averaged
velocities $\left< \bar{v}_x\right>$.  In (b) we show the
time-averaged velocities for another calculation with the same
parameters as in (a), but evaluating the momentum flux at $x=x_O$.
All variables are nondimensionalised with the LJ
potential units ($\sigma$ for length and $\tau=(\sigma^2
m/\epsilon)^{1/2}$ for time).
}
\end{figure}

\begin{figure}[h]
\includegraphics{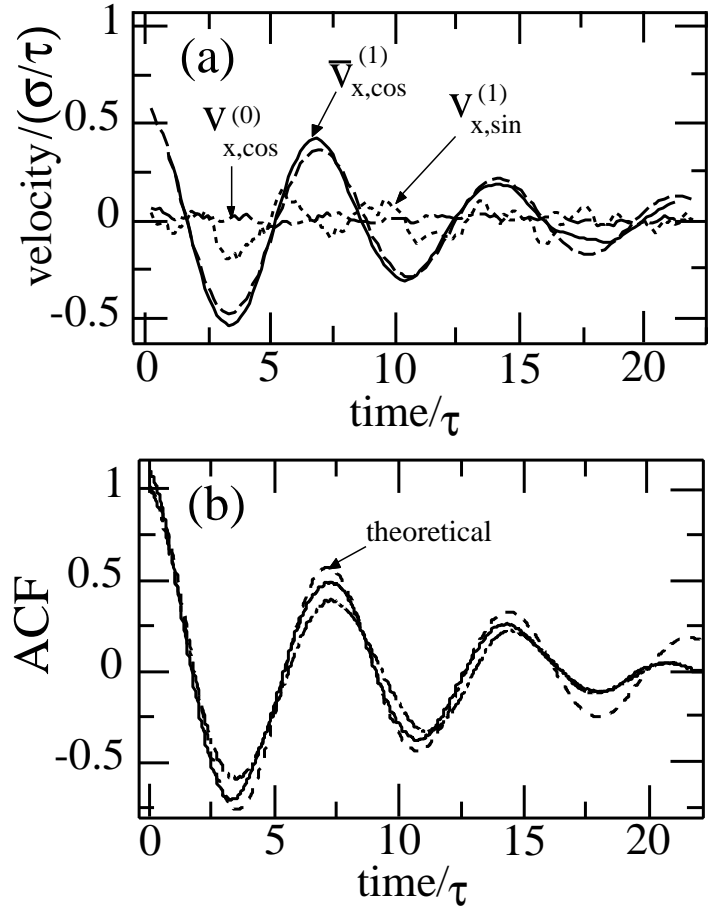}
\caption{(a) Time evolution of the Fourier transforms of
$x$-velocity $\bar{v}_x^{(n)}$ for the pure cosinusoidal perturbation
of Fig.  5.  In (b), the nondimensional autocorrelation function (ACF)
of $\bar{v}_x^{(1)}(t)$, corresponding to the pure cosinusoidal
perturbation (solid line) and to another initial perturbation with
$\left\{\bar{v}_{x,\cos}^{(1)}(0),\,\bar{v}_{x,\sin}^{(1)}(0)\right\}=\left\{0.60,\,0.25\right\}$
(dash-dotted line). In (a) and (b) the dashed line is the theoretical
hydrodynamic solution.  The remaining parameters are the same as those
in Fig. 5. Variables are nondimensionalised with the LJ potential
units ($\sigma$ for length and $\tau=(\sigma^2 m/\epsilon)^{1/2}$ for
time).
}
\end{figure}

\begin{figure}[h]
\includegraphics{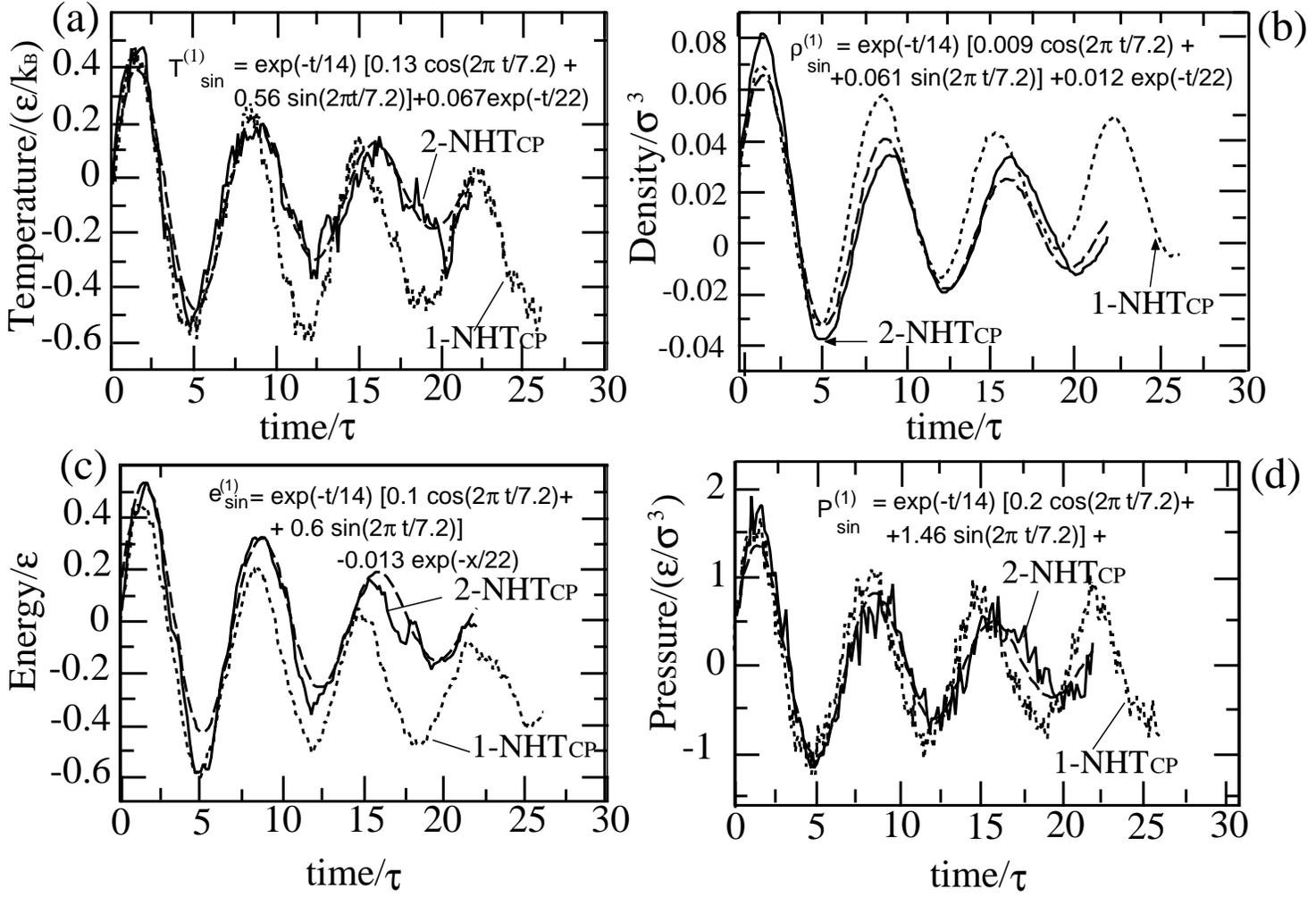}
\caption {Time dependence of the Fourier amplitudes of the 
of (a) temperature $\bar{T}_{\sin}^{(1)}$, (b) density
$\rho_{\sin}^{(1)}(=\Sigma_i \exp(ikx_i)/N$), (c) energy per particle
$\bar{e}_{\sin}^{(1)}$, and (d) pressure $\bar{P}_{\sin}^{(1)}$. All
quantities are nondimensionalised with the LJ potential units
($\sigma$ for length and $\epsilon$ for energy). Comparison is made
between a simulation with two Nos\'e-Hoover thermostats per
C$\rightarrow$P cell (2-NHT$_{CP}$), and another using only one
(1-NHT$_{CP}$, dotted line).  In each figure, the analytical
hydrodynamic expressions for the dominant Fourier component (in dashed
lines) are explicitly written. The initial amplitudes were
$\bar{v}_{x,\cos}^{(1)}(0)=0.60$, $\bar{T}_{\sin}^{(1)}(0)=-0.06$,
$\bar{P}_{\sin}^{(1)}(0)=0.25$, and $\rho_{\sin}^{(1)}(0)=0.022$.
}
\end{figure}

\begin{figure}[h]
\includegraphics{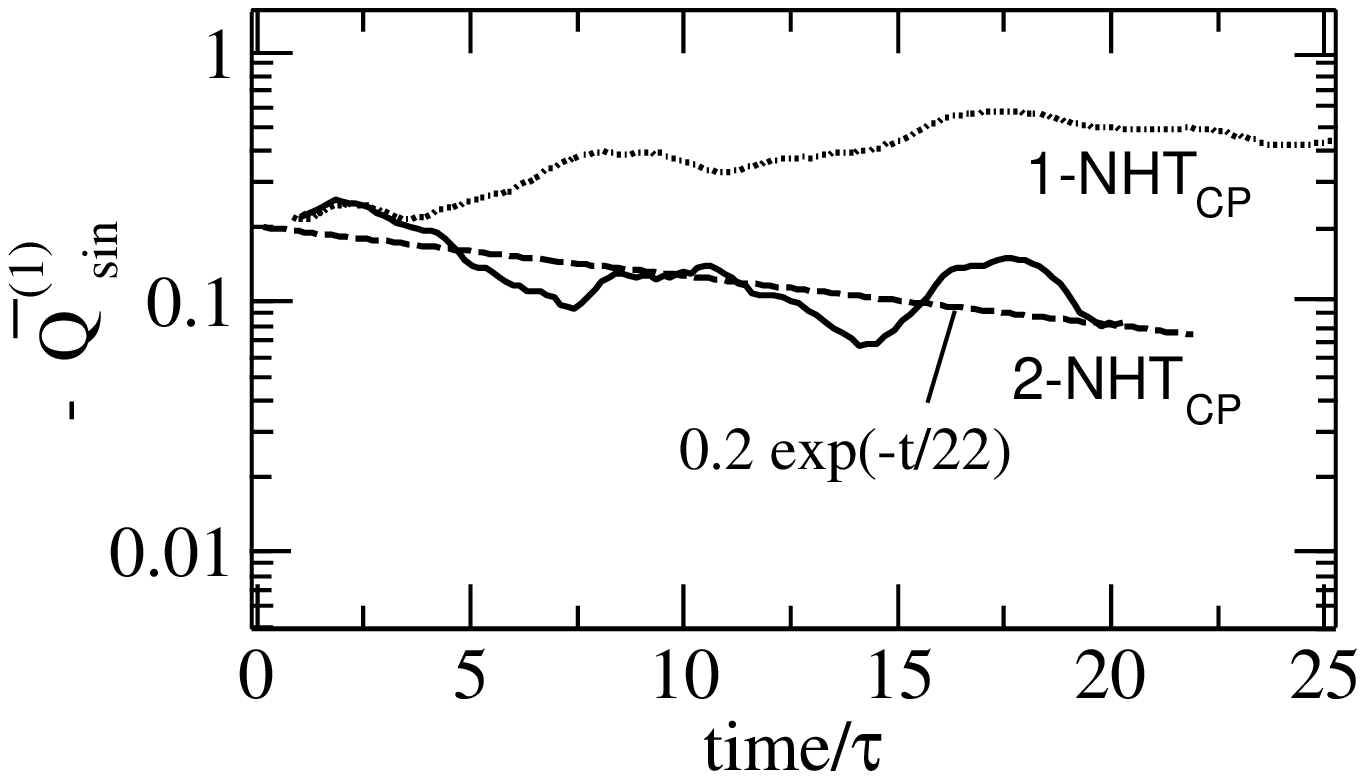}
\caption{The main Fourier amplitude of the heat density
perturbation $-\left<\bar{Q}_{\sin}^{(1)}\right>(t)$ time-averaged
along $\Delta t_C =1.0$ (and multiplied by $-1$).  The parameters are
the same as those of Fig. 5 (and 8) and heat density is 
in units of $\epsilon/\sigma^3$.  The dashed line corresponds to
the theoretical decay. Comparison is made between two (2-NHT$_{CP}$)
and one (1-NHT$_{CP}$) Nos\'e-Hoover thermostats per C$\rightarrow$P
cell.  The latter violates the second law of thermodynamics.
}
\end{figure}


\thebibliography{99}

\bibitem{Gar00} A. Garcia, J. Bell, Wm. Y. Crutchfield and B. Alder, 
J. Comp. Phys.  {\bf 154}, 134 (1999). 

\bibitem{Tho95} S. T. O'Connel and P. A. Thompson,  Phys. Rev. E, {\bf 52} R5792 (1995).

\bibitem{Had97} N. Hadjiconstantinou and A. Patera,
Int. J. Mod. Phys. C, {\bf 8}, 967 (1997).  {\bf 8}, 967 (1997); ibid.
 Phys. Rev. E {\bf 59}, 2, 2475 (1999).

\bibitem{Liao98} J. Li, D. Liao and  S. Yip, Phys. Rev. E
{\bf 57}, 6, 7259 (1998).

\bibitem{Flek00} E. G. Flekkoy, G. Wagner and J. Feder
Europhys. Lett. {\bf 52}(3), 271-276 (2000).

\bibitem{EOS} K. Johnson, J. A. Zollweg, and K. E. Gubbins,
Mol. Phys. {\bf 78} 591-618 (1993).

\bibitem{TRANSP} D. M. Heyes, Chem. Phys. Lett., {\bf 153} (4), 319 (1988)
\bibitem{BulkVis} P. Borgelt, C. Hoheisel and G. Stell, Phys. Rev. A {\bf 42}(2), 789 (1990).
\bibitem{Evans} D. J. Evans and G. P. Morris, {\it Statistical mechanics of Nonequilibrium Liquids} (Academic Press, London, 1990).

\bibitem{Frenkel.book} Frenkel and B. Smith, {\it Understanding molecular simulations} (Academic Press, London, 1996). 

\bibitem{INSERT} R. Delgado-Buscalioni and P. V. Coveney, submitted (2003).

\bibitem{note} After submission but
prior to publication of this paper we have implemented 
further improvement in the {\sc usher}
algorithm. The new {\sc usher} scheme, to appear in Ref.\cite{INSERT},
reduces the number of iterations to 8-30 for densities within the
range $\rho=0.5-0.8$. 

\bibitem{Ciccotti} C. Trozzi and G. Ciccotti, Phys. Rev. A, {\bf 29}, 916-925, (1984); A. Tenenbaum, G. Ciccotti, and Renato Gallico,  Phys. Rev. A {\bf 25} 2778 (1982).
\bibitem{HanMcD} J.P. Hansen and I.R. McDonald, {\it Theory of simple
liquids} (Academic Press, London, 1986).
\bibitem{TRITTON} D. J. Tritton, {\it Physical Fluid Dynamics} 
(Oxford Science Publications,  1988).
\end{document}